\documentclass[a4paper,10pt,fleqn]{article}
\pdfoutput=1
\usepackage{jheppub}
\usepackage[T1]{fontenc}
\usepackage{setspace}
\usepackage{enumitem}
\usepackage{amsmath}
\usepackage{amsthm}
\usepackage{bm}
\usepackage{mathtools}
\usepackage{caption}
\usepackage{slashed}
\usepackage{subcaption}
\usepackage{tikz}
\usepgflibrary{arrows}
\usetikzlibrary{calc}

\definecolor{jlab_red}{RGB}{192,39,45}
\definecolor{jlab_orange}{RGB}{249,102,0}
\definecolor{jlab_blue}{RGB}{47,122,121}
\definecolor{jlab_green}{RGB}{65,125,10}

\hypersetup{%
pdftitle = {HigherAngMomKtopipi},
pdfsubject = {QCD},
pdfauthor = {Hansen and Peterken},
colorlinks = {true},
filecolor = {black},
linkcolor = {jlab_blue},
menucolor = {black},
citecolor = {jlab_green},
urlcolor = {jlab_green},
}{}

\newcommand{\Sec}[0]{section}

\newcommand{\Eq}[0]{eq.}

\newcommand{\KSS}[0]{Kim:2005gf}

\newcommand{\Luscher}[0]{Luscher:1986n2,Luscher:1991n1}

\newcommand{\LL}[0]{Lellouch:2000pv}
\newcommand{\RG}[0]{Rummukainen:1995vs}

\newcommand{\AngMomContaminates}[0]{%
Meyer:2011um,%
Briceno:2014uqa}

\newcommand{\TwistBC}[0]{%
Bedaque:2004ax}

\newcommand{\Gparity}[0]{%
Christ:2019sah,
Kelly:2012eh,
Kim:2003xt}

\newcommand{\Truncation}[0]{%
\Luscher,%
\RG,%
\KSS}

\newcommand{\shortEandME}[0]{%
\Luscher,
\LL,
\RG,
\KSS,
Christ:2005gi,
Meyer:2011um,
Feng:2014gba,
Briceno:2014uqa}

\newcommand{\NoRescale}[0]{Briceno:2014uqa}

\newcommand{\GenericOneToTwo}[0]{%
\RG,%
\KSS,
Christ:2005gi,
Meyer:2011um,
Feng:2014gba,
Briceno:2014uqa,%
Briceno:2015csa}

\newcommand{\timelikecite}[0]{%
Feng:2014gba,
Andersen:2018mau,
Gerardin:2019rua
}

\newcommand{\pigpipicite}[0]{%
Briceno:2015dca,
Briceno:2016kkp,
Alexandrou:2018jbt
}

\newcommand{\InfVolLim}[0]{%
Lellouch:2000pv,%
Meyer:2011um,%
Hansen:2016qoz%
}

\title{Higher partial wave contamination in finite-volume 1-to-2 transitions }
\author[a]{M.~T.~Hansen}
\author[a]{and T.~Peterken}
\affiliation[a]{Higgs Centre for Theoretical Physics, School of Physics and Astronomy, The University of Edinburgh, Edinburgh EH9 3FD, UK}

\emailAdd{maxwell.hansen@ed.ac.uk}
\emailAdd{t.peterken@sms.ed.ac.uk}

\abstract{In their seminal work, Lellouch and L{\"u}scher derived a conversion factor relating a finite-volume matrix element, calculable using numerical lattice QCD, with the infinite-volume decay amplitude for $K \to \pi \pi$. The conversion factor depends on the $\pi \pi \to \pi \pi$ scattering amplitude with the same total isospin (either zero or two) as the $\pi \pi$ decay channel. Although an infinite tower of $\pi \pi \to \pi \pi$ partial-wave components affect the conversion factor, the $S$-wave ($\ell=0$) component is expected to dominate, and only this contribution is included in the well-known Lellouch-L{\"u}scher factor, with other $\pi \pi \to \pi \pi$ partial-wave amplitudes formally set to zero. However, as the precision of lattice calculations increases, it may become important to assess the systematic uncertainty arising from this approximation. With this motivation, we compare the $S$-wave-only results with those truncated at the next contaminating partial wave: the $G$-wave ($\ell=4$) for zero total momentum in the finite-volume frame and the $D$-wave ($\ell=2$) otherwise. Using the general framework for $1 \overset{\mathcal J}{\to} 2$ transitions derived in ref.~\cite{Briceno:2014uqa}, we quantify the effect of higher partial waves for systems with zero and non-zero total momentum as well as with anti-periodic boundary conditions, presenting both generic numerical examples and results for realistic $\pi \pi$ amplitudes taken from chiral perturbation theory and dispersive analysis. We also consider the accidental degeneracy occurring in the 8$^{\text {th}}$ excited state of the zero-momentum system. This exhibits qualitatively new features at $\ell=4$, not seen in the $\ell=0$ truncation.
}

\begin{document}

\maketitle
\flushbottom
\abovedisplayskip 11pt
\belowdisplayskip 11pt
\clearpage

\section{Introduction}

In the early 2000s, Lellouch and L{\"u}scher derived a formalism to extract $K \to \pi \pi$ decay amplitudes from finite-volume matrix elements calculable in lattice QCD \cite{Lellouch:2000pv}. This method has since been applied by the RBC-UKQCD collaboration (see e.g.~refs.~\cite{Bai:2015nea,Blum:2015ywa,Abbott:2020hxn}) in a first-principles calculation of the $K \to \pi \pi$ amplitudes, with both allowed final states (total isospin zero and two), leading to a first-principles understanding of the $\Delta I = 1/2$ rule and a determination of the CP violating parameter $\epsilon'/\epsilon$.

The original derivation of Lellouch and L{\"u}scher assumes vanishing spatial momentum in the finite-volume frame, and also that the effect of the $\ell=4$ (and higher) partial waves in $\pi\pi\to\pi\pi$ scattering can be neglected. Although the incoming kaon can only couple to an $S$-wave two-pion final state, neglecting the $\ell = 4$ component in $\pi \pi \to \pi \pi$ scattering amounts to an approximation in the value of the conversion factor relating the finite-volume matrix element to the infinite-volume decay amplitude. In addition, the original derivation only applies when the $\pi \pi$ finite-volume state used to construct the matrix element is below the 9$^{\text {th}}$ state (8$^{\text {th}}$ excited state) of the system. Because the final-state energy must match the incoming kaon mass, $m_K$, this translates to a constraint that the volume be sufficiently small for the 8$^{\text {th}}$ excited state to sit above $m_K$. As described in ref.~\cite{Lellouch:2000pv}, the subtlety arises because the 8$^{\text {th}}$ excited state exhibits an accidental degeneracy that invalidates the original derivation. As we show in this work, this is in fact closely tied to the effects of angular-momentum truncation.

Subsequent to ref.~\cite{Lellouch:2000pv}, a series of publications has lifted the original assumptions to provide a generic framework for extracting $0 \overset{\mathcal J}{\to} 2$ and $1 \overset{\mathcal J}{\to} 2$ transitions from finite-volume information~\cite{Lin:2001ek,Detmold:2004qn,Kim:2005gf,Christ:2005gi,Meyer:2011um,Hansen:2012tf,Briceno:2012yi,Bernard:2012bi,Agadjanov:2014kha,Briceno:2014uqa,Feng:2014gba,Briceno:2015csa}. The formalism at present holds for any number of two-particle channels, including any desired angular-momentum truncation. The more general formulas have been applied to study processes such as $\gamma^* \to \pi \pi$ \cite{\timelikecite}, $\pi \gamma \to \pi \pi$ \cite{\pigpipicite} and most recently $K \gamma \to K \pi$ \cite{Radhakrishnan:2022ubg}. In particular, ref.~\cite{Radhakrishnan:2022ubg} also discusses the role of $S$-wave and $P$-wave mixing in the finite-volume matrix elements. See also ref.~\cite{Briceno:2021xlc} for instructive details on the implementation of the $1 \overset{\mathcal J}{\to} 2$ formalism in the case of coupled two-particle channels in the final state. Generalisations for transitions with three-particle final states have also been recently derived in refs.~\cite{Hansen:2021ofl,Muller:2020wjo}.

This work uses the formalism of ref.~\cite{Briceno:2014uqa} in order to examine the effect of higher-partial-wave contamination in the context of $K \to \pi \pi$ decays, with many of the conclusions directly relevant to other $0 \overset{\mathcal J}{\to} 2$ and $1 \overset{\mathcal J}{\to} 2$ processes. As computing power increases and lattice calculations become more precise, the contamination from $\ell=4$ could become relevant. Similarly, with lattice volumes as they are at present, the kaon mass is safely below the 8$^{\text {th}}$ energy level. But again, the potential exists for studies with lattice volumes in the regime of $m_\pi L \approx 12$ (where $m_\pi$ is the pion mass and $L$ the periodicity of the finite, cubic, spatial volume) for which the accidentally degenerate state can play a role. (See, for example, ref.~\cite{Fritzsch:2021klm}.)

In ref.~\cite{Lin:2001ek}, Lin et al.~sketch an alternative derivation of the Lellouch-L{\"u}scher formalism and argue that the original result is in fact valid all the way up to the inelastic (four-pion) threshold, even when this is above the $8^{\text{th}}$ excited energy level. From the perspective of the general formalism of ref.~\cite{Briceno:2014uqa}, we can confirm this claim with a few caveats: First, while the $S$-wave quantization condition correctly gives the finite-volume energies when all higher partial waves identically vanish, this will never strictly hold for a realistic physical system. The scenario of validity for the $S$-wave trunction is therefore that the shift in finite-volume energies due to higher partial waves is below the statistical uncertainty of the calculation. In this case, the $S$-wave quantization misses one of the two distinct energy levels arising from the split of the accidentally degenerate level, and the corresponding Lellouch-L{\"u}scher factor is not predicted. At the same time, when higher partial waves are non-zero but statistically negligible, the naive Lellouch-L{\"u}scher factor is valid for one of the two states split from the accidental degeneracy, as claimed by ref.~\cite{Lin:2001ek}.

The remainder of this work is organised as follows:
In section~\ref{sec:fvformalism} we review the general formalism for finite-volume energies and matrix elements when the effects of partial wave mixing are included. In section~\ref{sec:exp_num} we present new analytic and numerical results based on the general formalism. In particular, in section~\ref{sec:swave_only} we review how the general relations can be used to recover the standard $S$-wave-truncated result of Lellouch and L{\"u}scher and in sections~\ref{sec:gwave} and \ref{sec:dwave} we show how corrections to the standard Lellouch-L{\"u}scher factor can be expressed perturbatively in terms of the relevant higher-partial-wave K-matrices, for zero and non-zero momentum in the finite-volume frame, and for twisted boundary conditions. We estimate the numerical size of such contaminations for a range of volumes making use of a scattering amplitude parametrized via a scattering length, with a range of values taken for the latter. In section~\ref{sec:ADs} we then consider effects that cannot be seen in the perturbative expansion, including resonance poles in the higher-partial-wave amplitudes as well as the effect of the accidentally degenerate state.

Section~\ref{sec:kpipi_realistic} then applies the results to physical $K\to\pi\pi$ decays, using $\pi \pi \to \pi \pi$ scattering amplitudes based on chiral perturbation theory as well as experimental-data-driven determinations. We find that the effect of partial wave mixing is at the sub-per-mille level for the case of zero total spatial momentum (with both periodic and anti-periodic boundary conditions). By contrast, the effect can become significant for non-zero total momentum in the finite-volume frame, due to mixing with the $\ell=2$ ($D$-wave) $\pi \pi \to \pi \pi$ amplitude. A specific example is the $M_\pi L \approx 6.0$ with one unit of total momentum. In this case, a finite-volume energy near $M_K$ arises for the $I=0$ channel and the corresponding Lellouch-L{\"u}scher factors has a $\sim 10 \%$ correction from the $D$-wave amplitude.

We conclude in section~\ref{sec:conc} and also include three appendices collecting technical details used in the main text.

\section{Finite-volume formalism}
\label{sec:fvformalism}

In this section we review aspects of the formalism derived in refs.~\cite{\shortEandME}, focusing on the results relevant to this work. We restrict attention to a single channel of two spin-zero particles with degenerate mass. We specifically have in mind the isospin-two and isospin-zero two-pion channels in iso-symmetric QCD. As described in the introduction, our main aim is to explore the contamination, from higher angular momenta, in finite-volume matrix elements used to calculate $K \to \pi \pi$. The section is broken into two subsections: In \Sec~\ref{sec:energies} we review L{\"u}scher's relation between finite-volume energies and the elastic $\pi \pi \to \pi \pi$ scattering amplitude and in \Sec~\ref{sec:MEs} we review the formula of Lellouch and L{\"u}scher, and its subsequent extensions, describing finite volume effects in $1 \to 2$ transition matrix elements.

\subsection{Energies}
\label{sec:energies}

Consider a relativistic quantum field theory, confined to a finite, periodic, cubic spatial volume with extent $L$. The centre-of-mass energy is chosen such that only a single channel of degenerate spinless particles contributes. In particular, this is realized in iso-symmetric QCD by selecting the internal quantum numbers of a two-pion state and requiring $E_{\sf{cm}}< 4m$ where $m$ is the physical pion mass. Then, as was derived in refs.~\cite{\Luscher}, the discrete finite-volume energy spectrum is given (up to neglected, exponentially suppressed finite-$L$ corrections) by the roots of the function
\begin{equation}
\det_{\Gamma} \big[ M^{[\Gamma]}(E, \boldsymbol P, L) \big ] = 0 \,,
\label{eq:DeltaDef}
\end{equation}
where
\begin{equation}
M^{[\Gamma]}(E, \boldsymbol P, L) \equiv
\mathbb P^{[\boldsymbol P]}_{\Gamma } \cdot
\Big [ \widetilde {\mathcal K}(E_{\sf{cm}}) + \widetilde F(E, \boldsymbol P, L)^{-1} \Big ]
\cdot
\mathbb P^{[\boldsymbol P]}_{\Gamma }\,,
\label{eq:Mdef}
\end{equation}
and the determinant runs over a set of angular-momentum indices, discussed in more detail below. The label $\Gamma$ indicates that the matrix is projected to an irreducible representation (irrep) of the finite-volume symmetry group.

The expressions here depend on the energy, $E$, and spatial-momentum, $\boldsymbol P$, both defined in the finite-volume frame. The total momentum satisfies $\boldsymbol P = (2 \pi/L) \boldsymbol d$ where $\boldsymbol d$ is a three-vector of integers, $\boldsymbol d \in \mathbb Z^3$. Below we also use the shortand $\boldsymbol P = [d_x d_y d_z]$. The K-matrix, $\widetilde {\mathcal K}$, depends only on the CMF energy, defined in the usual way
\begin{equation}
E_{\sf{cm}} \equiv \sqrt{E^2 - \boldsymbol P^2} \,.
\end{equation}
The quantities $ \mathbb P^{[\boldsymbol P]}_{\Gamma }$, $\widetilde {\mathcal K}(E_{\sf{cm}})$ and $\widetilde F(E, \boldsymbol P, L)$ are each matrices in angular momentum space and carry two sets of spherical harmonic indices, $\ell m$.

We use tildes on $\widetilde {\mathcal K}$ and $\widetilde F$ as these are re-scaled, relative to the definitions appearing, for example, in ref.~\cite{\NoRescale}, by factors of the CMF momentum
\begin{equation}
p_{\sf{cm}}^2 \equiv E_{\sf{cm}}^2/4-m^2 \,.
\end{equation}
In particular we define
\begin{equation}
\label{eq:angularmompartialwave}
\widetilde {\mathcal K}_{\ell' m', \ell m}(E_{\sf{cm}}) \equiv \delta_{\ell' \ell} \delta_{m' m} \frac{1}{p_{\sf{cm}}^{2 \ell}} \mathcal K^{(\ell)}(E_{\sf{cm}}) \,, \\
\end{equation}
where the object on the right-hand side is given by
\begin{equation}
\label{eq:k_matrix_param}
\mathcal K^{(\ell)}(E_{\sf{cm}}) \equiv \frac{16 \pi E_{\sf{cm}}}{p_{\sf{cm}}} \tan \delta_{\ell}(E_{\sf{cm}}) \,,
\end{equation}
with $\delta_{\ell}(E_{\sf{cm}})$ denoting the standard scattering phase shift in the $\ell$th partial wave.
This re-scaling has the advantage that $\widetilde {\mathcal K}$ starts at order $(p_{\sf{cm}})^0$ in a small momentum expansion, for all $\ell$.

The remaining factors
appearing in eq.~\eqref{eq:DeltaDef}, $\widetilde F$ and $\mathbb P^{[\boldsymbol P]}_{\Gamma }$,
are known kinematic quantities:

$\widetilde F(E, \boldsymbol P, L)$ is a matrix with both on- and off-diagonal entries, encoding the effects of the finite volume. We give the explicit definition used here in appendix~\ref{app:Ffuncs}, which includes factors of $(p_{\sf{cm}})^{\ell}$, as compared to the analogous functions in ref.~\cite{\NoRescale}, in order to match the definition of $\widetilde {\mathcal K}$. In this work we require the explicit expressions only up to $\ell = 4$, for $\boldsymbol P = \boldsymbol 0$, and up to $\ell = 2$, for $\boldsymbol P\neq \boldsymbol 0$. In the following section we also consider the effect of twisted boundary conditions. Like non-zero total momentum, this is encoded via a straightforward modification to the definition of the finite-volume matrix \cite{\TwistBC}.

$\mathbb P^{[\boldsymbol P]}_{\Gamma }$ is a projector matrix that restricts eq.~\eqref{eq:DeltaDef} to the target quantum numbers. It carries the total momentum label, since the value of $\boldsymbol P$ dictates the symmetry group of the system. In particular, for vanishing spatial momentum, the relevant group is the octahedral group, $O_h^P$, including parity. This has 10 irreps, including the trivial irrep denoted $A_{1g}$ where $g$ indicates parity $+$. For nonzero momenta, the symmetry group is reduced to the subgroup of $O_h^P$ leaving $\boldsymbol P$ invariant, referred to as the little group or point group of $\boldsymbol P$. In this case parity is no longer a good quantum number and the trivial irrep is denoted $A_1$.

The matrices $\widetilde {\mathcal K}(E_{\sf{cm}})$, $\widetilde F(E, \boldsymbol P, L)$ and $\mathbb P_{\Gamma}^{[\boldsymbol P]}$ are each formally infinite-dimensional. Thus, to use eq.~\eqref{eq:DeltaDef} in practice, one must truncate to a finite subspace. As is discussed, e.g. in refs.~\cite{\Truncation}, if one approximates $\mathcal K^{(\ell)} = 0$ for $\ell > \ell_{\sf{max}}$ then $\widetilde F$ and $\mathbb P_{\Lambda}^{[\boldsymbol P]}$ can be truncated to the same subspace without further approximation.
The most extreme choice is given by $\ell_{\sf{max}} = 0$,
in which case eq.~\eqref{eq:Mdef} becomes
\begin{equation}
M^{[A_{1}]}(E, \boldsymbol P, L) \equiv
\mathcal K^{(0)}(E_{\sf{cm}}) + \overline F_{00}(E, \boldsymbol P, L)^{-1} \,,
\end{equation}
where $\overline F_{00} = \widetilde F_{00,00}$ is a shorthand that will be expanded to the other entries of $\widetilde F$ in the following paragraph. The left-hand side is labeled with the superscript $[A_1]$ which is a shorthand for $A_{1g}$ in the zero-momentum case and $A_1$ otherwise.

To go beyond this simple truncation one needs to choose the value of $\boldsymbol P$. In the case of $\boldsymbol P = \boldsymbol 0$ the lowest lying off-diagonal elements of $\widetilde F$ are those with indices $(4m,00)$ and $(00,4m)$ with $m=-4,0,4$. Setting $\ell_{\sf{max}} = 4$, one finds
\begin{equation}
\label{eq:MA1gexplicit}
M^{[A_{1g}]}(E, L) \equiv \begin{pmatrix} {\mathcal K}^{(0)}(E) & 0 \\ 0 & {\mathcal K}^{(4)}(E)/p_{\sf{cm}}^8 \end{pmatrix}+
\begin{pmatrix}
\overline F_{00}(E,L) & \overline F_{04}(E,L) \\ \overline F_{40}(E,L) & \overline F_{44}(E,L)
\end{pmatrix}^{-1} \,,
\end{equation}
where we take the shorthand of omitting the $\boldsymbol P = \boldsymbol 0$ label and dropping the ${\sf{cm}}$ subscript when $E = E_{\sf{cm}}$ (also using $p = p_{\sf{cm}}$). Here the projector to $A_{1g}$ allows one to compress to a two-dimensional subspace and remove the $m$ index. For example
\begin{equation}
\overline F_{44}(E,L) \equiv P_m^{[A_{1g}]} \, \widetilde F_{4m,4m'}(E, L) \, P_{m'}^{[A_{1g}]} \,,
\end{equation}
where $P_{4}^{[A_{1g}]} = P_{-4}^{[A_{1g}]} = \frac12 \sqrt{\frac{5}{6}}$, $P_{0}^{[A_{1g}]} = \frac12 \sqrt{\frac{7}{3}}$ and all remaining entries vanish. This also gives our general meaning of $\overline F_{\ell' \ell}$, the simplified matrix entries resulting from projection of $\widetilde F$ to the trivial irrep.

In the case of $\boldsymbol P\neq \boldsymbol 0$ the $S$-wave couples to the $D$-wave and we set $\ell_{\text{max}} = 2$ to write
\begin{equation}
M^{[A_{1}]}(E, \boldsymbol P, L) \equiv \begin{pmatrix} {\mathcal K}^{(0)}(E_{\sf{cm}}) & 0 \\ 0 & {\mathcal K}^{(2)}(E_{\sf{cm}})/p^4_{\sf{cm}} \end{pmatrix}
+
\begin{pmatrix}
\overline F_{00}(E,\boldsymbol P,L) & \overline F_{02}(E,\boldsymbol P,L) \\ \overline F_{20}(E,\boldsymbol P,L) & \overline F_{22}(E,\boldsymbol P,L)
\end{pmatrix}^{-1} \,.
\end{equation}
In appendix \ref{app:Ffuncs} we give explicit expressions for the entries of $\overline F$ required in the following sections.

An analogous modification to boosting the system is to apply twisted boundary conditions, for which the single-particle fields satisfy
\begin{align}
\phi(\boldsymbol{x}+L \hat {\boldsymbol e}_i)=e^{i\theta_i}\phi(\boldsymbol{x}) \,,
\end{align}
for $i \in \{x,y,z\}$. This can be naturally package into a three-vector, $\boldsymbol{\theta} = (\theta_x, \theta_y, \theta_z)$, and has the consequence of modifying the discrete set of finite-volume momenta to
\begin{align}
\boldsymbol k=(2\pi/L)( \boldsymbol{n} + \boldsymbol{\theta} ) \,,
\end{align}
with $\boldsymbol n \in \mathbb Z^3$. We only consider twisted systems for which the two-particle states are periodic so that $\boldsymbol P \in (2 \pi/L) \mathbb Z^3$.

In this work we restrict attention to anti-periodic boundary conditions, $\boldsymbol{\theta}=(1/2,1/2,1/2)$, and only consider the effect for $\boldsymbol P = \boldsymbol 0$. For this special case the underlying symmetry group is the full octahedral group, as in the periodic case, and $\ell = 4$ gives the lowest lying non-zero contamination to $A_{1g}$. As a result one can again use eq.~\eqref{eq:MA1gexplicit}, up to a straightforward modification of the finite-volume function
\begin{align}
\overline F_{\ell \ell'}(E, L) \to \overline F^{\boldsymbol \theta} _{\ell \ell'}(E, L) \,,
\end{align}
where $\overline F^{\boldsymbol \theta}$ is defined in appendix \ref{app:Ffuncs}. We stress that, for the zero-isospin ($I=0$) channel of two pions in QCD, it is not obvious that the quark and gluon boundary conditions can be designed to achieve the twisting described here. This has however been resolved in refs.~\cite{\Gparity} through the advent of G-parity boundary conditions. Though various subtleties arise, the key point for this work is that $\overline F^{\boldsymbol \theta}$ predicts the correct low-lying finite-volume energies, also in the G-parity case.

This completes our review of the equations governing finite-volume energies for $\ell_{\sf{max}} = 0, 2$ or $4$. We turn now to the corresponding relation on matrix elements in the context of the $K \to \pi \pi$ weak decay.

\subsection{Matrix elements}
\label{sec:MEs}

The formalism for extracting the $K \to \pi \pi$ decay amplitude from a finite-volume matrix element, for both isospin zero and isospin two final states, was derived by Lellouch and L{\"u}scher in ref.~\cite{Lellouch:2000pv}. The result was originally given for the case of $\boldsymbol P = \boldsymbol 0$ and with the $\ell_{\sf{max}} = 0$ assumption applied at the beginning of the derivation. This has since been extended to a general formalism for 1-to-2 ($1 \overset{\mathcal J}{\to} 2$) transition amplitudes mediated by generic local currents in refs.~\cite{\GenericOneToTwo}.

The purpose of the present work is to explore the consequences of the general formalism, specifically the result of ref.~\cite{Briceno:2014uqa}, for higher angular-momentum components in $K \to \pi \pi$.
We turn immediately to the key relation
\begin{equation}
\label{eq:LLref}
\big \vert \langle E_{\pi \pi}, \pi \pi, \text{out} \vert \mathcal H(0) \vert K \rangle \big \vert^2 \Big \vert_{E_{\pi \pi} = E^{\sf{cm}}_n(L)} =2 m_K L^6 \, \mathcal C \big (E_{n}(L), L \big ) \, \big \vert \langle E_n, \boldsymbol P, L , A_1 \vert \mathcal H(0) \vert K, \boldsymbol P, L \rangle \big \vert^2 \,.
\end{equation}
Here the left-hand side is the magnitude-squared infinite-volume matrix element of the weak Hamiltonian density between the single kaon state $\vert K \rangle$ and the $S$-wave two-pion asymptotic state. This becomes the decay rate (up to the missing phase space factor) when $E_{\pi \pi} = m_K$, with $m_K$ giving the kaon mass, such that the Hamiltonian density does not inject energy. However, the relation as written holds (up to neglected exponentially suppressed $L$-dependence) for any value of $E_{\pi \pi} < 4 m$. Note that the infinite-volume matrix element is evaluated at the CMF finite-volume energy
\begin{equation}
E_n^{\sf{cm}}(L) = \sqrt{E_n(L)^2 - \boldsymbol P^2} \,.
\end{equation}

The right hand side of eq.~\eqref{eq:LLref} shows that the observable can be determined from a finite-volume matrix element of the same Hamiltonian density, between $\vert K, \boldsymbol P, L \rangle$, the finite-volume ground state with one unit of strangeness or anti-strangeness and spin zero (more precisely in the $A_{1,u}$ or $A_2$ irrep) and $\langle E_n, \boldsymbol P, L, A_1 \vert$, a finite-volume state with two-pion quantum numbers. We work throughout with finite-volume states normalised to unity. In addition to the kinematic factors in eq.~\eqref{eq:LLref}, one finds the non-trivial proportionality factor~\cite{Briceno:2014uqa,Lellouch:2000pv}
\begin{equation}
\label{eq:Cdef}
\mathcal C \big (E_{n}(L), \boldsymbol P, L \big ) \equiv \frac{\cos^2 \delta_0 (E_{\sf{cm}}) }{\text{adj} \big [ M^{[A_1]}(E, \boldsymbol P, L) \big ]_{00}} \,
\frac{\partial \det \! \big [ M^{[A_1]}(E, \boldsymbol P, L) \big ] }{\partial E}
\bigg \vert_{E = E_n(L)} \,.
\end{equation}
Here $\text{adj}$ is the adjugate of the matrix, defined as
\begin{equation}
\text{adj}X = X^{-1} \det X \,.
\end{equation}

A key emphasis of this work, already recognized in refs.~\cite{\AngMomContaminates}, is that $ \mathcal C \big (E_{n}(L), \boldsymbol P, L \big )$ receives corrections from an infinite tower of $\ell$ values. As explained in the introduction, this follows intuitively from the fact that $\langle E_n, \boldsymbol P, L, A_1 \vert$ is normalized to unity and itself receives contributes from an infinite tower of $\ell$. Thus the size of its $\ell = 0$ component must depend on the contamination from $\ell > 0$.
This effect is quantified by the appearance of both the determinant and the matrix adjugate in eq.~\eqref{eq:Cdef}.

Note the factor of $ \cos^2 \delta_0(E)$ appearing in eq.~\eqref{eq:Cdef}. This arises because we define $M^{[A_1]}$ with the K-matrix rather than the scattering amplitude. As a result, this version of the conversion factor gives the infinite-volume amplitude with the real part of the strong phase, $ \text{Re} [e^{i \delta_0(E)}] = \cos \delta_0(E)$, divided out. We correct this in the definition of $\mathcal C$ such that the left-hand side of eq.~\eqref{eq:LLref} is the usual infinite-volume matrix element.

We close here with a few words on the exact meaning of the two-pion finite-volume energy $E_n(L)$. Our perspective is that this quantity is determined in the lattice calculation and therefore satisfies the quantization condition, up to neglected $e^{- m L}$ corrections, with $\ell_{\sf{max}}$ set to an arbitrarily large value.
In other words, if the system dynamics and the precision of the calculation are such that the higher angular momenta become important, than $E_n(L)$ must be understood to contain such contributions.

\section{Expansions and numerical results}
\label{sec:exp_num}

We turn now to details of the formalism reviewed in the previous section for specific values of $\ell_{\sf max}$, the total momentum $\boldsymbol P$ and the twist angle $\boldsymbol \theta$, and the values of the K-matrices governing the two-particle interactions. We break this discussion into five sub-sections. The first four of these illustrate some of the general phenomenology of these relations while the final subsection gives specific results for the $K \to \pi \pi$ decay, based on $\pi \pi \to \pi \pi$ scattering phase shifts constrained by experiment, dispersive methods and low-energy effective theory.

\subsection{\texorpdfstring{$\ell_{\sf{max}}=0$}{S-wave truncation}}
\label{sec:swave_only}

As a warm up, here we recover the usual Lellouch-L{\"u}scher factor, derived by setting $\ell_{\sf{max}} = 0$ within $M^{[A_1]}$. In this case $\det M^{[A_1]} = M^{[A_1]}$ and $\text{adj} M^{[A_1]} = 1$, implying
\begin{align}
\label{eq:Cmanip1}
\mathcal C(E_n(L), \boldsymbol P, L) & = \big [\! \cos^2 \delta_0 (E_{\sf{cm}}) \big ] \Big [ \partial_E \widetilde F_{00}(E,\boldsymbol P,L)^{-1}+\partial_E \mathcal K^{(0)} (E_{\sf{cm}}) \Big ] \bigg \vert_{E = E_n(L)} \,, \\[5pt]
& = \big [\! \cos^2 \delta_0 (E_{\sf{cm}}) \big ] \frac{\partial}{\partial E} \frac{16 \pi E_{\sf{cm}}}{p_{\sf{cm}}} \big [\tan \phi(E, \boldsymbol P, L) + \tan \delta_{0}(E_{\sf{cm}}) \big ] \bigg \vert_{E = E_n(L)} \,,
\label{eq:fullC0}
\end{align}
where we have introduced
\begin{equation}
\overline F_{00}(E,\boldsymbol P,L)^{-1} \equiv \frac{16 \pi E_{\sf{cm}}}{p_{\sf{cm}}}\tan \phi(E,\boldsymbol P,L) \,, \qquad \qquad \mathcal K^{(0)}(E_{\sf{cm}}) = \frac{16 \pi E_{\sf{cm}}}{p_{\sf{cm}}} \tan \delta_{0}(E_{\sf{cm}}) \,.
\label{eq:Cmanip4}
\end{equation}
Note that ${\partial_E}$ and ${\partial_{E_{\sf{cm}}}}$ are both defined with $\boldsymbol{P}$ held constant and hence
\begin{align}
\frac{\partial}{\partial E}= \frac{\partial \sqrt{E^2 - \boldsymbol P^2}}{\partial E} \frac{\partial}{\partial E_{\sf{cm}}} = \gamma \frac{\partial}{\partial E_{\sf{cm}}} \,,
\end{align}
where $\gamma = E/E_{\sf cm}$ is the usual boost factor.
One can show that \Eq~\eqref{eq:Cmanip4} can be written as
\begin{align}
\label{eq:CisDplusLL}
\mathcal C(E_n(L), \boldsymbol P, L) = \mathcal C^{[\ell_{\sf{max}}=0]}(E_n(L), \boldsymbol P, L) + \mathcal D(E_n(L), \boldsymbol P, L) \,,
\end{align}
where
\begin{align}
\mathcal C^{[\ell_{\sf{max}}=0]}(E, \boldsymbol P, L) & \equiv \frac{4 \pi \gamma E_{\sf{cm}}^2}{p^2_{\sf{cm}}} \frac{\partial}{\partial p_{\sf{cm}}} [ \phi(p_{\sf{cm}}, \boldsymbol P, L) + \delta_{0}(p_{\sf{cm}})] \,, \\
\begin{split}
\mathcal D(E, \boldsymbol P, L) & \equiv - \frac{16 \pi \gamma m^2}{p^3_{\sf cm}} [\cos^2\delta_0(E_{\sf cm})] \left[\tan \phi(E, \boldsymbol P, L) + \tan \delta_{0}(E_{\sf{cm}})\right]
\\
& + \frac{16 \pi \gamma E_{{\sf cm}}}{p_{{\sf cm}}} \bigg [\frac{\cos^2\delta_0(E_{\sf cm})}{\cos^2 \phi(E, \boldsymbol P, L)} - 1 \bigg ] \frac{\partial \phi(E, \boldsymbol P, L)}{\partial E_{{\sf cm}}} \,.
\end{split}
\end{align}
Here, $\mathcal C^{[\ell_{\sf{max}}=0]}$ corresponds to the factor derived in ref.~\cite{Lellouch:2000pv}, extended to non-zero total momentum in the finite-volume frame. The second factor, $\mathcal D(E, \boldsymbol P, L)$, vanishes identically when evaluated at any $E_n(L)$ that exactly satisfies the $\ell_{\sf{max}} = 0$ quantization condition (both lines can be shown to vanish separately). However, $\mathcal D$ can give a non-zero contribution for less aggressive truncations.

For the special case of $\boldsymbol P = 0$, $\phi$ only depends on the combination $q = p_{\sf{cm}} L/(2 \pi)= p L/(2 \pi)$ and one can write
\begin{align}
\mathcal C^{[\ell_{\sf{max}} = 0]}(E_n(L), L) & = \frac{4 \pi E^2}{p^3} \bigg [ q \frac{\partial}{\partial q} \phi(q) + p \frac{\partial}{\partial p} \delta_{0}(p) \bigg ] \bigg \vert_{E = E_n(L)} \,,
\end{align}
where we have suppressed the $\boldsymbol P = \boldsymbol 0$ label.
Combining this with eq.~\eqref{eq:LLref} and setting $E = m_K$, one exactly recovers the main result of Lellouch and L\"uscher, eq.~(4.5) of ref.~\cite{Lellouch:2000pv}.

\subsection{\texorpdfstring{$\ell_{\sf{max}} = 4$}{G-wave truncation}, zero momentum}
\label{sec:gwave}

We now examine the corrections to $\mathcal C(E_n(L), L)$, in the $\boldsymbol P = \boldsymbol 0$ case, arising from the $\ell = 4$ contamination to the $A_{1g}$ state. To do so, we explicitly evaluate eq.~\eqref{eq:Cdef} for the $2 \times 2$ matrix defined in eq.~\eqref{eq:MA1gexplicit}. As already anticipated above, throughout this section
$E_n(L)$ is defined as the solution to the $\ell_{\sf{max}} = 4$ quantization condition.

We begin by defining the difference
\begin{equation}
\label{eq:first_delta_def}
\Delta(E_n(L), L) \equiv \frac{\mathcal C(E_n(L), L) - \mathcal C^{[\ell_{\sf max} = 0]}(E_n(L), L)}{\mathcal C^{[\ell_{\sf max} = 0]}(E_n(L), L)} \,.
\end{equation}

Adding zero in a useful form, this can be written as
\begin{multline}
\label{eq:CleverAddZero}
\Delta(E_n(L), L) =\frac{1}{\mathcal C^{[\ell_{\sf max} = 0]}(E, L)} \bigg [ \mathcal{C}(E, L) -\cos^2 \delta_0 (E) \Big ( \partial_E \overline F_{00}(E,L)^{-1}+\partial_E \mathcal K^{(0)} (E) \Big )
\\[3pt]
+
\cos^2 \delta_0 (E)\Big ( \partial_E \overline F_{00}(E,L)^{-1}+\partial_E \mathcal K^{(0)} (E) \Big )- \mathcal C^{[\ell_{\sf{max}} = 0]}(E, L)\bigg ] \bigg \vert_{E = E_n(L)}
\,.
\end{multline}
The first line here represents the difference between the $\ell=4$ and $\ell=0$ truncation and the second line is the extra contribution from $\mathcal{D}$ which is now no longer zero as the energy levels deviate slightly from those predicted by the $\ell=0$ quantization condition.

To reach our final result for this subsection, we express all quantities in terms of $\overline F_{\ell' \ell}(E, L)$ and $\mathcal K^{(\ell)}(E)$. We then use the $\det M^{[A_1]} = 0$, truncated at $\ell=4$, to remove all explicit dependence on
$\mathcal K^{(0)}$. Finally, expanding to leading order in $\mathcal K^{(4)}$, also counting $\partial_E \mathcal K^{(0)}$ as suppressed relative to $\partial_E \overline F_{00}$, we reach
\begin{multline}
\Delta(E, L)
= \bigg [ (2m)^9 \frac{\partial}{\partial E} \frac{\mathcal K^{(4)}(E) }{p^8} \bigg ] \Delta^{[\partial \mathcal K(4)]}(E, L)
+ \bigg [ (2m)^8 \frac{\mathcal K^{(4)}(E) }{p^8} \bigg ] \Delta^{[ \mathcal K(4)]}(E,L)
\\
+ \mathcal O \big [ (\mathcal K^{(4)})^2, \mathcal K^{(4)} \partial_E \mathcal K^{(0)}/ \partial_E \overline F_{00} \big ] \,,
\label{eq:DeltaEexpanded}
\end{multline}
with
\begin{align}
\Delta^{[\mathcal K(4)]}(E,L) & \equiv \frac{1}{(2m)^8} \frac{1}{\partial_E \overline F_{00}(E, L)^{-1}}
\frac{\overline F_{04}(E,L)^2}{\overline F_{00}(E, L)^2} \Bigg [ 2
\frac{\partial_E \overline F_{04}(E,L)}{\overline F_{04}(E,L)} - 2 \frac{\partial_E \overline F_{00}(E,L) }{\overline F_{00}(E, L) } \\
& \hspace{40pt} + \frac{m^2}{E p^2} - \frac{2}{E p^2} \frac{1}{{\overline F}_{00}(E,L)} \frac{m^2 {\overline F}_{00}(E,L) - E p^2 \partial_E {\overline F}_{00}(E,L)}{1 + \big [ 16 \pi E {\overline F}_{00}(E,L)/p \big ]^2} \bigg ] \,, \\[10pt]
\Delta^{[\partial \mathcal K(4)]}(E,L) & \equiv \frac{1}{(2m)^9} \frac{1}{\partial_E \overline F_{00}(E, L)^{-1}} \frac{\overline F_{04}(E,L)^2}{\overline F_{00}(E,L)^2} \,.
\end{align}
The functions $\Delta^{[\mathcal K(4)]}(E,L) $ and $ \Delta^{[\partial \mathcal K(4)]}(E,L) $ only have physical meaning when evaluated at a solution, $E_n(L)$. The aim of eq.~\eqref{eq:DeltaEexpanded} is to express the higher-partial wave contamination in terms of kinematic functions and dynamic $\mathcal K^{(4)}$-dependence coefficients. We have also rescaled the latter by the appropriate powers of $(2m)/p$ such that the quantities in square brackets do not have barrier factor suppression.\footnote{The choice to work with $(2m)/p$ rather than $m/p$ here is motivated by a study of the one-loop $t$-channel exchange diagram that arises, for example, in $\lambda \phi^4$ theory. Decomposing this diagram in partial waves, one finds that factoring out $(2 m/p)^\ell$ rather than $ (m/p)^{\ell}$ leads to order-one coefficients and is thus representative of the dimensionless parameter governing the angular-momentum suppression.}

\begin{figure}[p]
\centering
\includegraphics[width=0.9\textwidth]{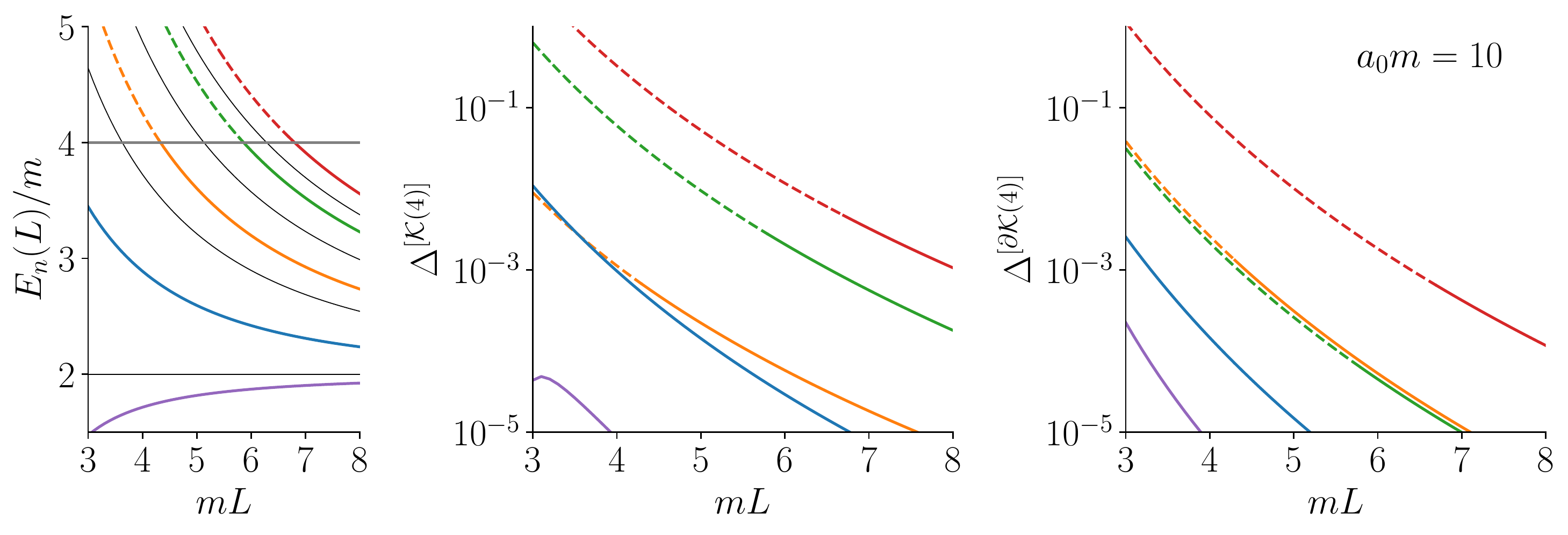}
\includegraphics[width=0.9\textwidth]{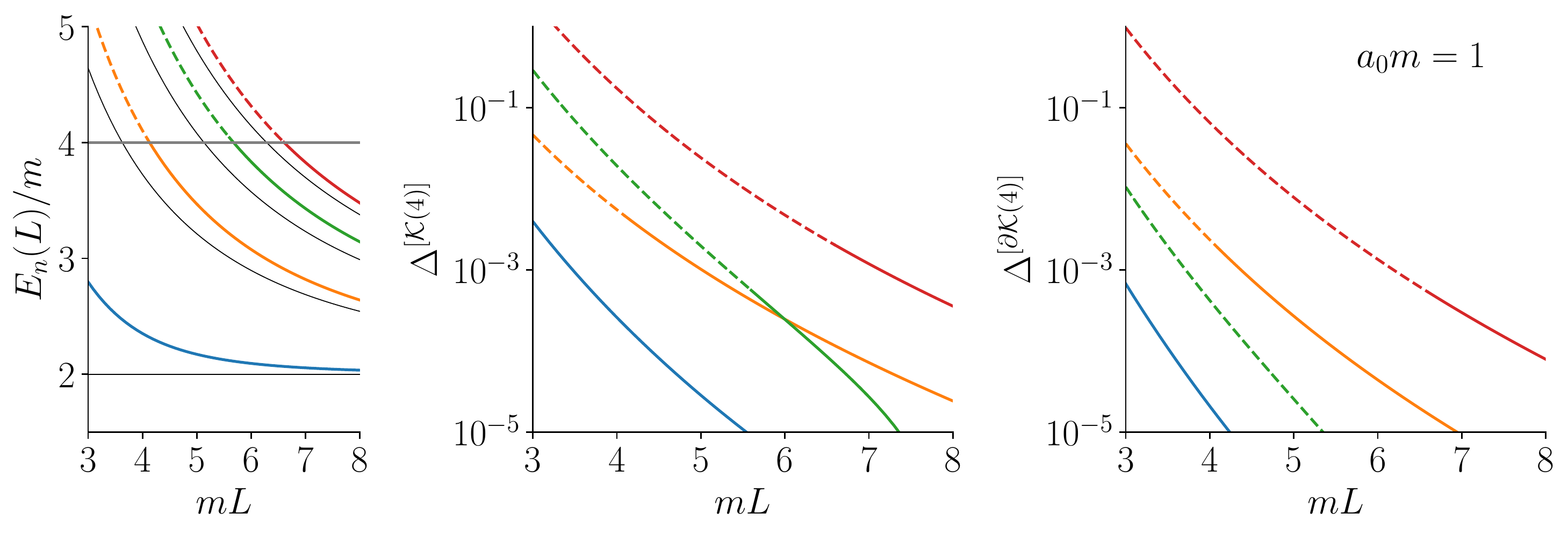}
\includegraphics[width=0.9\textwidth]{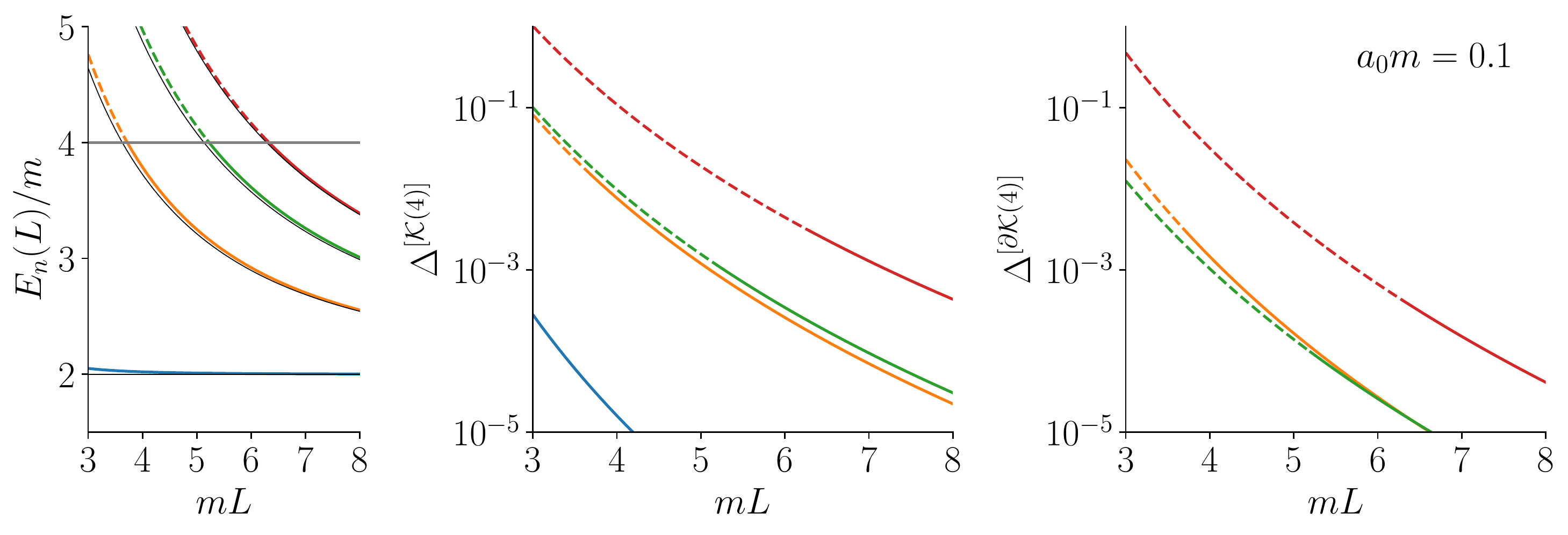}
\includegraphics[width=0.9\textwidth]{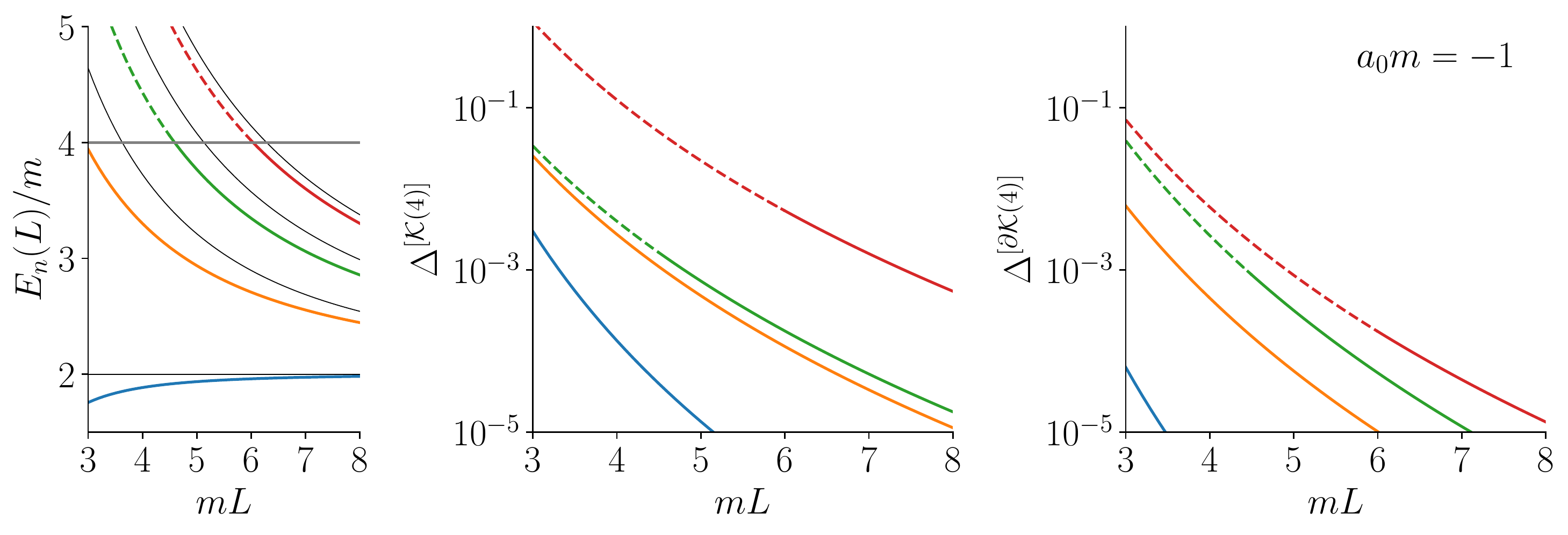}
\caption{$E_n(L)/m$, $\Delta^{[\mathcal K(4)]}(E_n(L), L)$ and $\Delta^{[\partial \mathcal K(4)]}(E_n(L), L)$ plotted verse $m L$ for various choices of the $S$-wave scattering length $m a_0$. All plots are for the trivial irrep $A_{1g}$, with vanishing total momentum $\boldsymbol P=[000]$ and periodic boundary conditions. The dashed segments indicate that $E_n(L) > 4 m$, implying that the formalism has neglected systematic uncertainties due to on-shell four-particle states. In the case of $ma_0=10$, a shallow bound state arises, resulting in the lowest lying finite-volume energy shown in magenta.}
\label{fig:P000_periodic}
\end{figure}

\begin{figure}[p]
\centering
\includegraphics[width=0.9\textwidth]{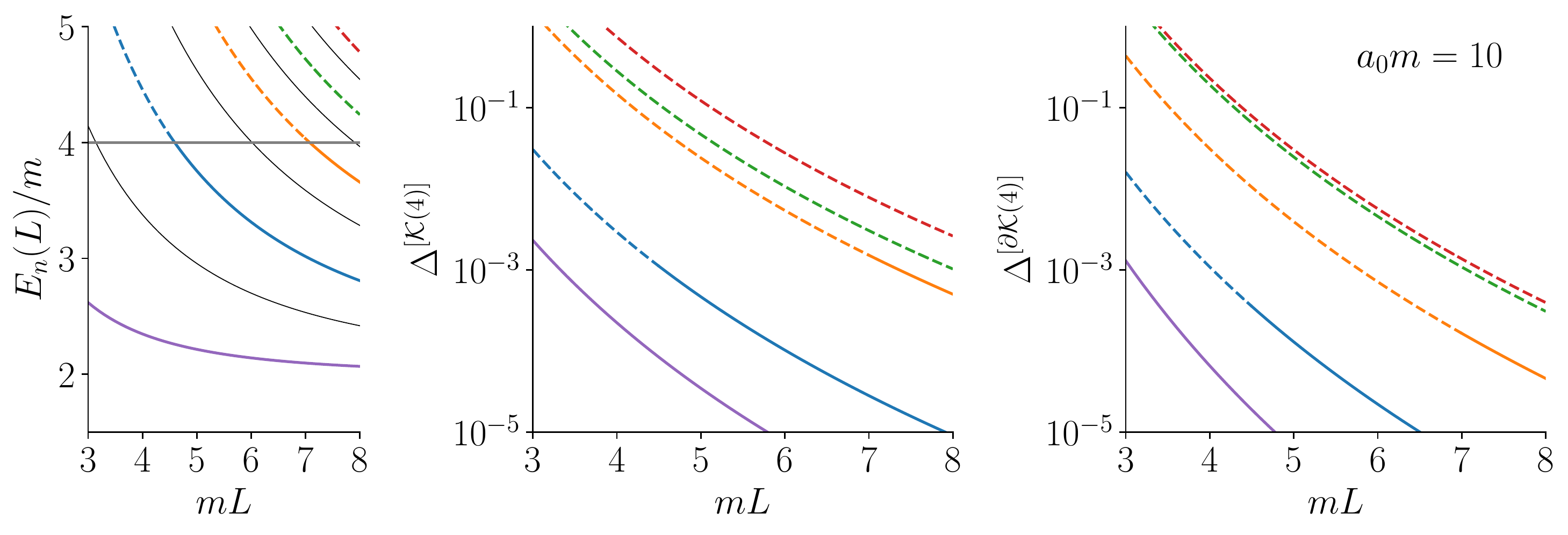}
\includegraphics[width=0.9\textwidth]{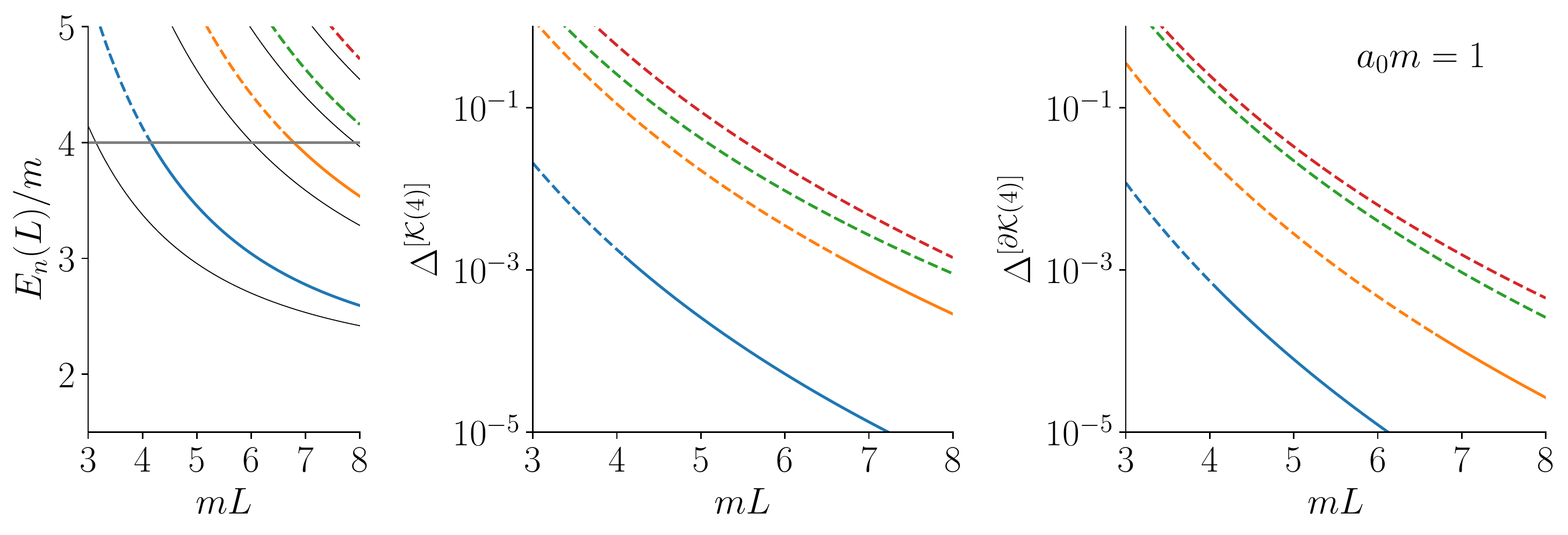}
\includegraphics[width=0.9\textwidth]{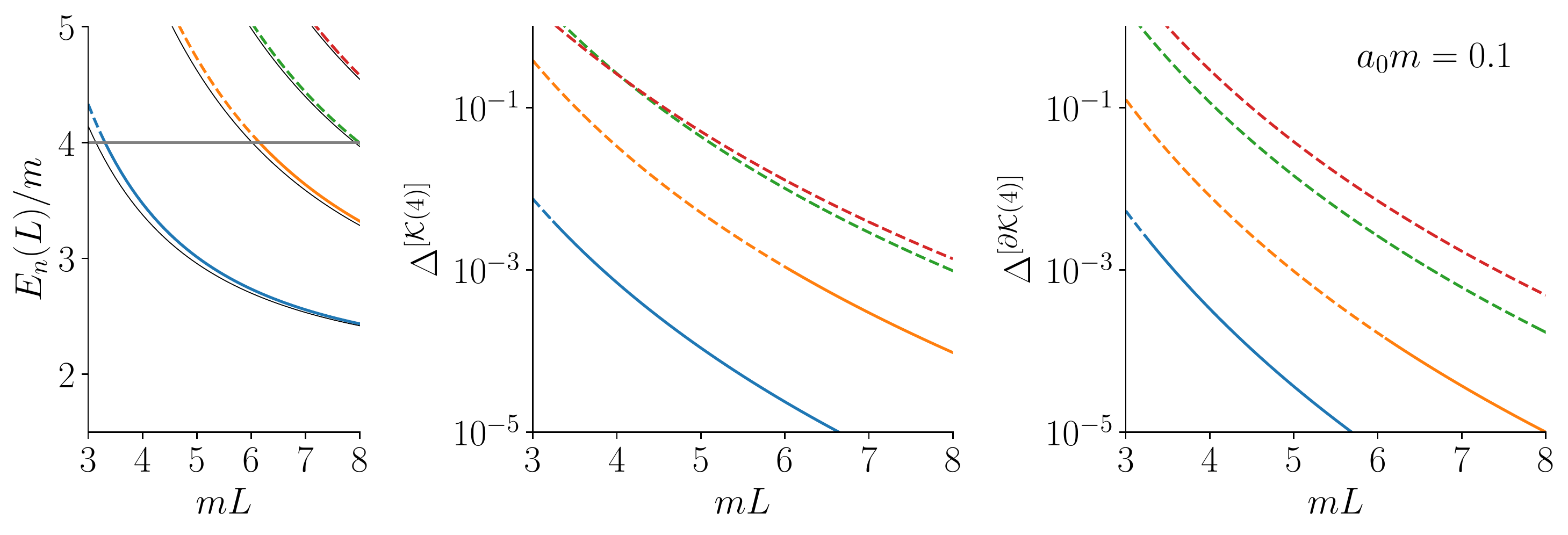}
\includegraphics[width=0.9\textwidth]{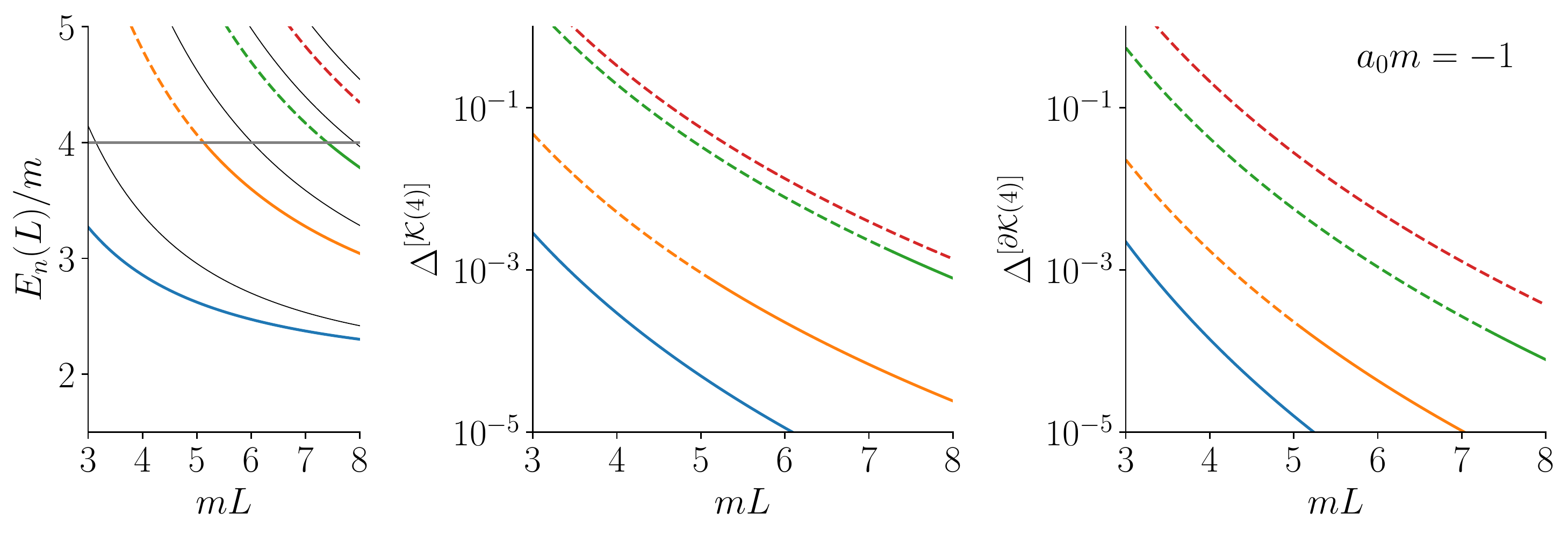}
\caption{As in figure~\ref{fig:P000_periodic} but with anti-periodic boundary conditions corresponding to twist vector $\boldsymbol \theta = (1/2,1/2,1/2)$.
}
\label{fig:P000_antiperiodic}
\end{figure}

Note that, while we take $E_n(L)$ to formally be the solution of the $\ell_{\sf{max}} = 4$ quantization condition, the effect of the higher partial wave on the finite-volume energy within $\Delta^{[\mathcal K(4)]}(E_n(L),L) $ contributes beyond the order that we control in the expansion of eq.~\eqref{eq:DeltaEexpanded}. It is therefore sufficient to take an $S$-wave only description to quantify the size of the coefficients $\Delta^{[\mathcal K(4)]}$ and $\Delta^{[\partial \mathcal K(4)]}$.
In the leftmost panel of figure \ref{fig:P000_periodic}, we plot the finite-volume energies
determined assuming the $S$-wave two-particle K-matrix is well described by the scattering length only
\begin{equation}
\label{eq:K0scatteringlengthonly}
\mathcal K^{(0)}(E_{\sf {cm}}) = - 16 \pi E_{\sf{cm}} a_0 \,,
\end{equation}
with various values of $m a_0$ as indicated by the plot labels. In the middle and right panels we then plot $\Delta^{[\mathcal K(4)]}(E_n(L),L) $ and $\Delta^{[\partial \mathcal K(4)]}(E_n(L),L) $, respectively, evaluated at these solutions. In figure~\ref{fig:P000_antiperiodic} we give the same quantities for the case of anti-periodic boundary conditions, corresponding to a twist of $\boldsymbol \theta = (1/2,1/2,1/2)$ on the two-pion state.

One generically finds that
the two coefficient functions
vary over many orders of magnitude for reasonable parameter choices, with the highest values corresponding to a ten percent correction. However, we will see that in the context of $K \to \pi \pi$ that the effect of higher angular momenta is highly suppressed, both by the finite-volume expressions entering the formalism and by the smallness of $\mathcal K^{(4)}(E_{\sf {cm}}) (2m/p)^8 $, i.e.~a suppression of higher angular-momentum components going beyond the usually claimed barrier factor suppression.

\subsection{\texorpdfstring{$\ell_{\sf{max}} = 2$, $\boldsymbol P=[001]$}{D-wave truncation, one unit of momentum}}
\label{sec:dwave}

\begin{figure}[p]
\centering
\includegraphics[width=0.9\textwidth]{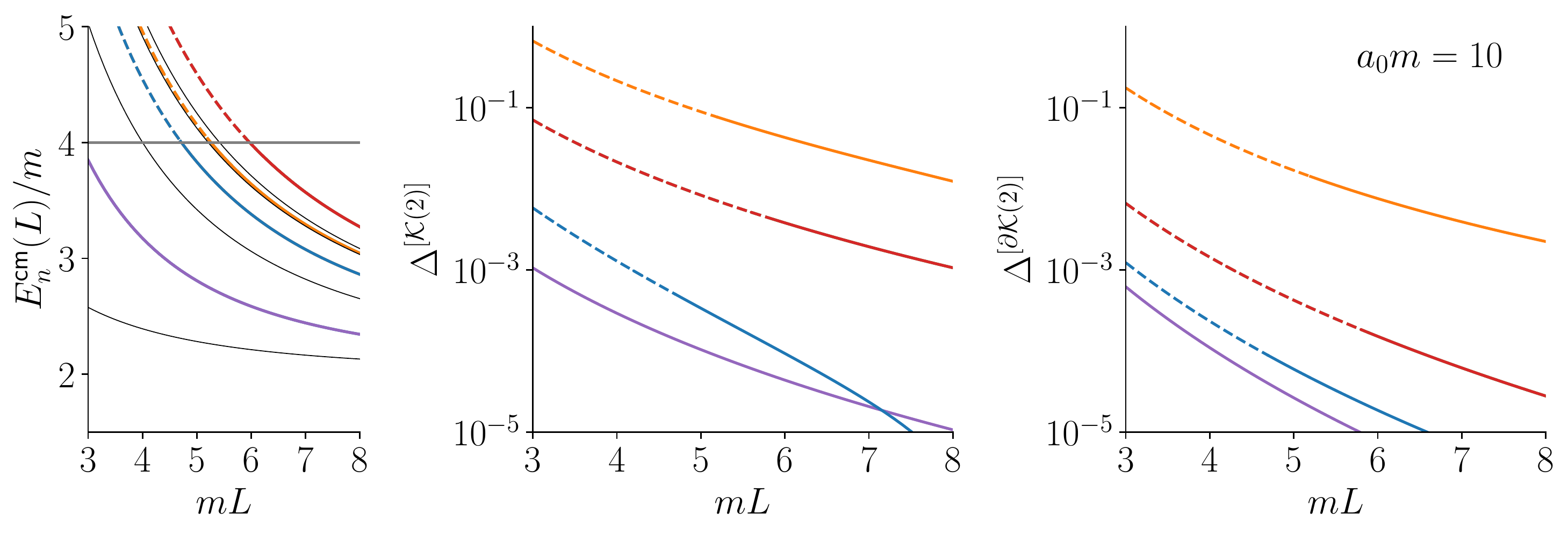}
\includegraphics[width=0.9\textwidth]{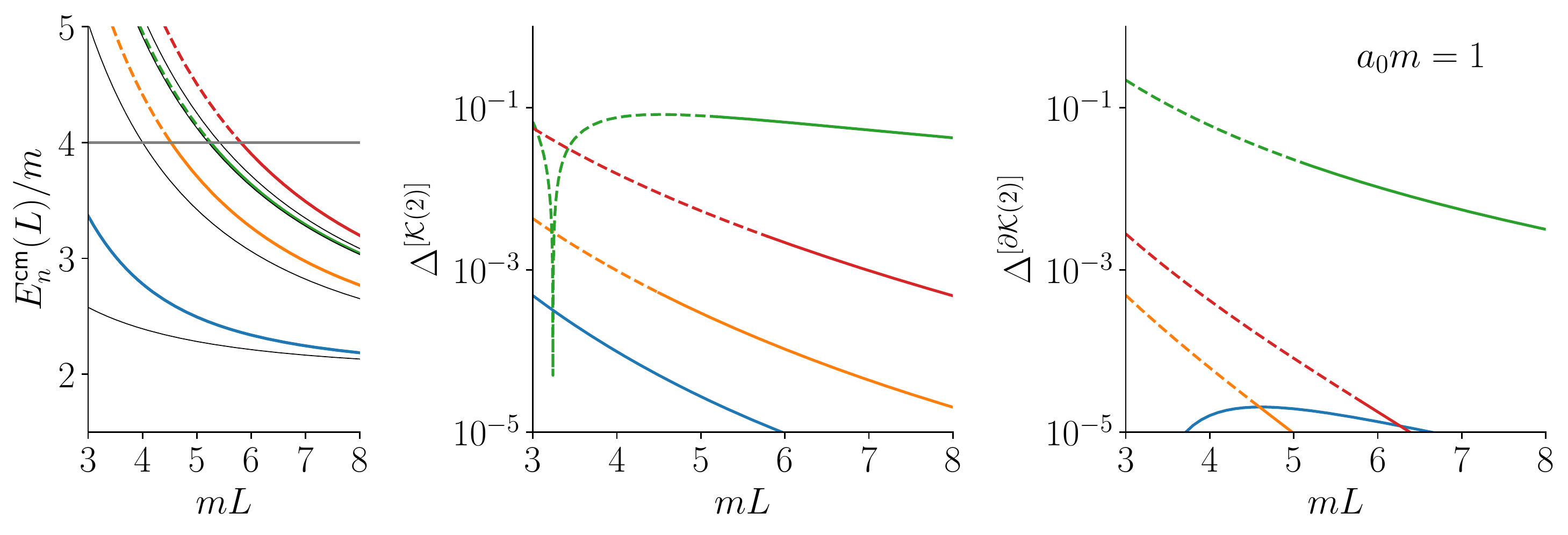}
\includegraphics[width=0.9\textwidth]{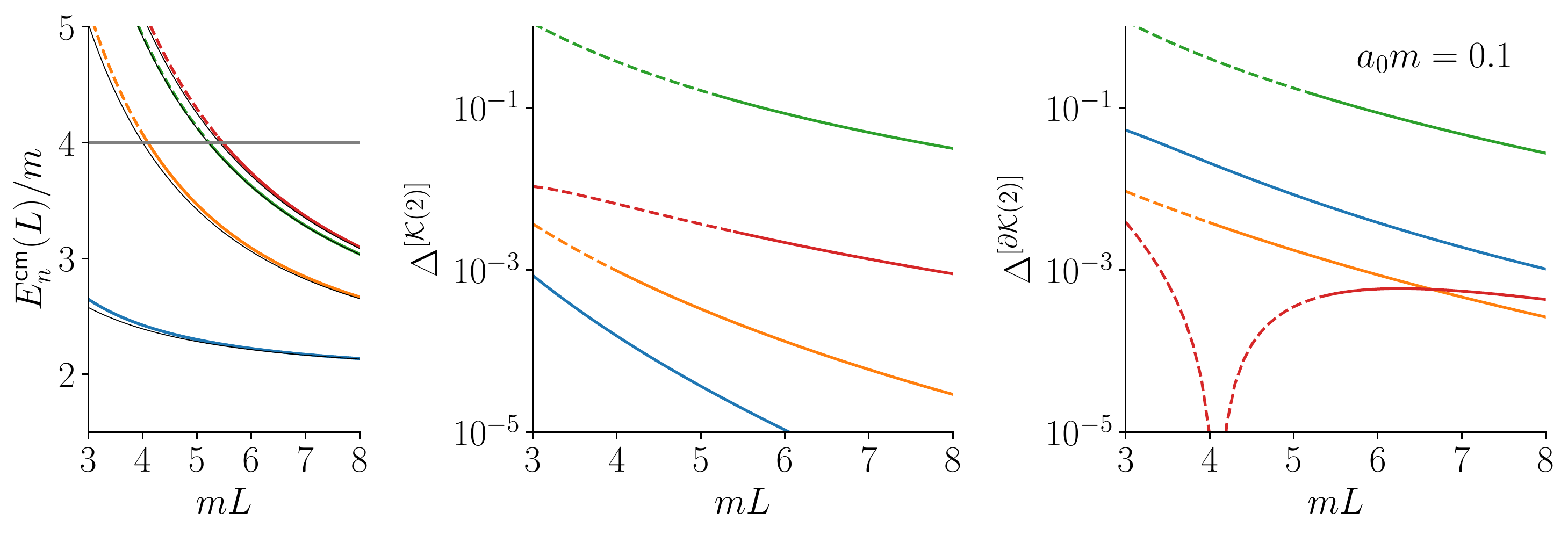}
\includegraphics[width=0.9\textwidth]{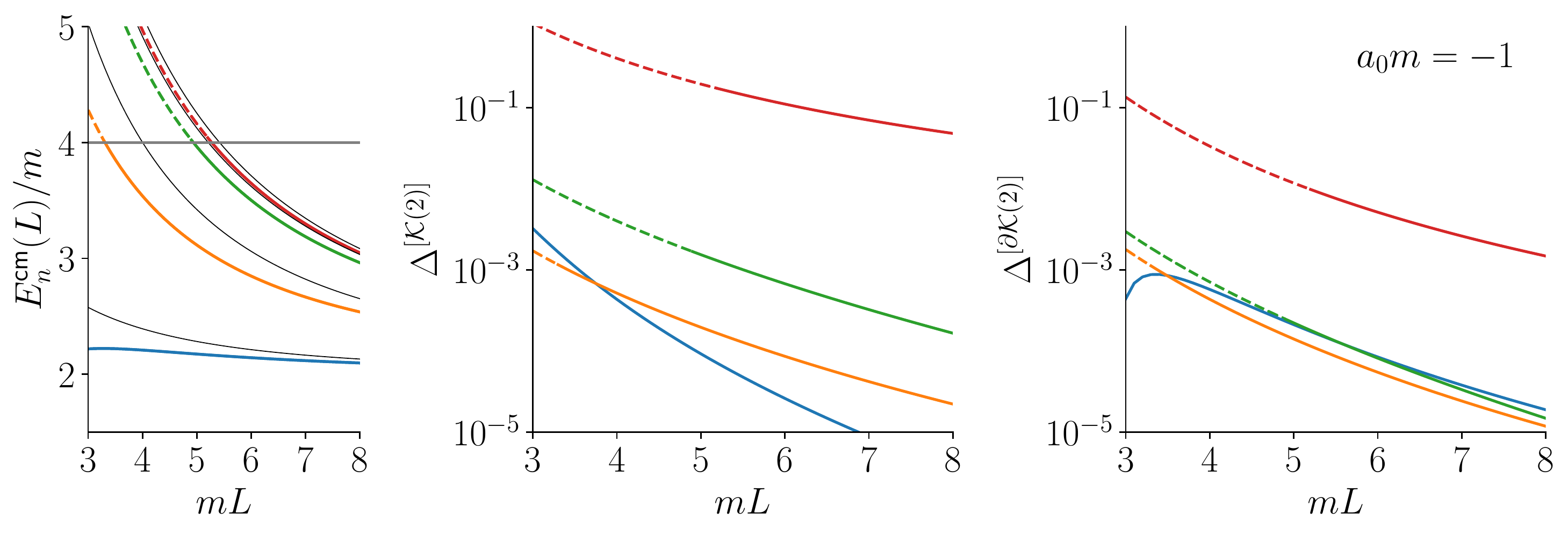}
\caption{$E^{\sf cm}_n(L)/m$, $\Delta^{[\mathcal K(2)]}(E_n(L), L)$ and $\Delta^{[\partial \mathcal K(2)]}(E_n(L), L)$ plotted verse $m L$ for various choices of the $S$-wave scattering length $m a_0$. All plots are for the trivial irrep $A_{1}$, with $\boldsymbol P=[001]$ and periodic boundary conditions. The dashed segments indicate that $E^{\sf cm}_n(L) > 4 m$, implying that the formalism has neglected systematic uncertainties due to on-shell four-particle states. In the case of $ma_0=10$, a shallow bound state arises, resulting in the lowest lying finite-volume energy shown in magenta.}
\label{fig:P001_periodic}
\end{figure}

We now turn to the case of non-zero total momentum in the finite-volume frame. In this case $\ell = 2$ (rather than $\ell = 4$) is the lowest non-zero partial wave contaminating the trivial finite-volume irrep. Setting $\ell_{\sf{max}} = 2$ then leads to the moving frame analog of $\Delta(E, L)$, which takes a very similar form to eq.~\eqref{eq:DeltaEexpanded} above:
\begin{multline}
\hspace{-15pt} \Delta(E, \boldsymbol P, L)
=
\bigg [ (2m)^5 \frac{\partial}{\partial E_{\sf{cm}}} \frac{\mathcal K^{(2)}(E_{\sf{cm}}) }{p^4_{\sf{cm}}} \bigg ] \Delta^{[\partial \mathcal K(2)]}(E, \boldsymbol P, L)
+
\bigg [ (2m)^4 \frac{\mathcal K^{(2)}(E_{\sf{cm}}) }{p_{\sf{cm}}^4} \bigg ] \Delta^{[ \mathcal K(2)]}(E,\boldsymbol P,L)
\\
+
\mathcal O \big [ (\mathcal K^{(2)})^2, \mathcal K^{(2)} \partial_E \mathcal K^{(0)}/\partial_E \overline F_{00} \big ] \,,
\label{eq:DeltaExpandedNonzeroP}
\end{multline}
with
\begin{align}
\begin{split}
\hspace{-15pt} \Delta^{[\mathcal K(2)]}(E,\boldsymbol P,L) & \equiv \frac{1}{(2m)^4}
\frac{1}{\partial_E \overline F_{00}(E, \boldsymbol P,L)^{-1}}
\frac{\overline F_{02}(E, \boldsymbol P,L)^2}{\overline F_{00}(E, \boldsymbol P,L)^2} \\
& \hspace{00pt} \times \bigg [ 2
\frac{\partial_E \overline F_{02}(E, \boldsymbol P,L)}{\overline F_{02}(E, \boldsymbol P,L)} - 2 \frac{\partial_E \overline F_{00}(E, \boldsymbol P,L) }{\overline F_{00}(E, \boldsymbol P,L) } + \gamma \frac{m^2}{E_{\sf cm} p_{\sf cm}^2} \\
& \hspace{5pt} - \frac{2}{E_{\sf cm} p_{\sf cm}^2} \frac{1}{{\overline F}_{00}(E, \boldsymbol P,L)} \frac{\gamma m^2 {\overline F}_{00}(E, \boldsymbol P,L) - E_{\sf cm} p_{\sf cm}^2 \partial_E {\overline F}_{00}(E, \boldsymbol P,L)}{1 + \big [ 16 \pi E_{\sf cm} {\overline F}_{00}(E, \boldsymbol P,L)/p_{\sf cm} \big ]^2} \bigg ] \,, \end{split} \\[10pt]
\hspace{-15pt} \Delta^{[\partial \mathcal K(2)]}(E,\boldsymbol P,L) & \equiv \frac{1}{(2m)^5} \frac{\gamma}{\partial_{E} \overline F_{00}(E,\boldsymbol P, L)^{-1}} \frac{\overline F_{02}(E,\boldsymbol P,L)^2}{\overline F_{00}(E,\boldsymbol P,L)^2} \,,
\end{align}
where $\gamma = E/E_{\sf cm}$. In figure~\ref{fig:P001_periodic} we plot the finite-volume energies as well as $\Delta^{[\mathcal K(2)]}(E_n(L), L)$ and $\Delta^{[\partial \mathcal K(2)]}(E_n(L), L)$, for various choices of the $S$-wave scattering length.

\newpage

\subsection{Beyond perturbative expansions}
\label{sec:ADs}

In the preceding subsections, we have described the effects of higher partial waves in cases where these cause a perturbative shift on the $S$-wave Lellouch-L{\"u}scher factors. In this sub-section, we turn to effects going beyond perturbation expansions that are captured by the general formalism. These arise for two distinct reasons: (i) the splitting of accidentally degenerate states and (ii) the breakdown in expansions when the magnitude of the higher-angular-momentum K-matrix is too large, including the appearance of resonances in higher angular-momentum channels.

Concerning the case of accidental degeneracy, we recall that the original Lellouch-L{\"u}scher relation for $K\to\pi \pi$ \cite{Lellouch:2000pv}, for systems with zero spatial momentum in the finite-volume frame, is valid for the ground state and for the first seven excited states. In the non-interacting limit, these eight lowest-lying states correspond to two identical particles with back-to-back momentum taking on squared magnitude values from 0 to 8 with, 7 omitted, in units of $(2 \pi)^2/L^2$. A subtlety arises for the 8$^{\text {th}}$ excited state, for which the dimensionless back-to-back momentum has a squared magnitude of $9$. This can be achieved with two types of integer vectors: $(1,2,2)$ and $(0,0,3)$, that are not related by octahedral transformations. The situation is referred to as an \textit{accidental degeneracy}. While the truncation with $\ell_{\sf max} = 0$ preserves this degeneracy, it is broken by including $\ell = 4$.

To study this case we return to the general quantization condition but now keep the full effects of the $\ell=4$ partial wave (restricting attention to $\boldsymbol P =[000]$ throughout this subsection). This corresponds to the condition that the matrix defined in eq.~\eqref{eq:MA1gexplicit} has vanishing determinant
\begin{equation}
\det \bigg [ \begin{pmatrix} {\mathcal K}^{(0)}(E) & 0 \\ 0 & {\mathcal K}^{(4)}(E)/p^8 \end{pmatrix}+
\begin{pmatrix}
\overline F_{00}(E,L) & \overline F_{04}(E,L) \\ \overline F_{40}(E,L) & \overline F_{44}(E,L)
\end{pmatrix}^{-1} \bigg ] = 0 \,.
\end{equation}

To provide numerical predictions, we must parametrize ${\mathcal K}^{(4)}(E)$. We do so in two different ways:

First, in direct analogy to our approach for the $S$-wave component, eq.~\eqref{eq:K0scatteringlengthonly}, we take the form of a leading order threshold expansion
\begin{equation}
\label{eq:g_wave_thr_param}
{\mathcal K}_{\sf thr}^{(4)}(E) = - 16 \pi E p^8 a_4^{\sf thr} \,.
\end{equation}
A subtlety with this parametrization is that it generates poles in the scattering amplitude on both Riemann sheets. In terms of the momentum coordinate, the pole condition can be written as
\begin{equation}
\big ( p_{\sf pole} \big )^9 = \frac{i}{a_4^{\sf thr}} \,,
\end{equation}
where solutions for which $\text{Im} \, p_{\sf pole} > 0$ are on the physical Riemann sheet and those for which $\text{Im} \, p_{\sf pole} < 0$ are on the unphysical sheet.
Strictly this implies that the parametrization is unphysical, since poles on the the physical sheet violate the causality of the theory.

Here we take a less strict perspective, knowing that any parameterization of the K-matrix will not describe the entire complex plane. We instead only require that the pole nearest to the scattering axis, primarily driving the behavior of $\mathcal K^{(4)}_{\sf thr}(E)$ for physical energies, is on the unphysical Riemann sheet as is required for a resonance pole.
This means that $a_4^{\sf{thr}} > 0$ is disallowed, as this implies that the nearest pole on the physical sheet, at
\begin{equation}
s_{\sf pole} = 4 \vert a_4^{\sf thr} \vert^{2/9} e^{i \pi/9} + 4 m^2 \,, \qquad \qquad \ \, (a_4^{\sf{thr}} > 0) \,, \ \text{physical\ sheet} \Longrightarrow \text{disallowed}\,,
\end{equation}
whearas $a_4^{\sf{thr}} < 0$ gives an acceptable description, with the nearest pole on the unphysical (second) Riemann sheet:
\begin{equation}
s_{\sf pole} = 4 \vert a_4^{\sf thr} \vert^{2/9} e^{-i \pi/9} + 4 m^2 \,, \qquad \qquad (a_4^{\sf{thr}} < 0) \,, \ \text{unphysical\ sheet} \Longrightarrow \text{allowed}\,.
\end{equation}
We have also investigated solving the quantization condition with both choices and find that $a_4^{\sf{thr}} < 0$ generates a physical finite-volume spectrum while $a_4^{\sf{thr}} > 0$ does not. The physical spectrum shows reduced $L$-dependence in the resonant region, which is expected since, in the limit of the resonance width going to zero, one approaches a decoupled state with exponentially suppressed volume dependence. The unphysical energies, by contrast, show \emph{increased} volume dependence in the region of the pole.

Additionally, as the threshold expansion is only motivated for low values of momenta, we consider an alternative parameterization:
\begin{equation}
\label{eq:K4_const_param}
{\mathcal K}_{\sf const}^{(4)}(E)=-32\pi a_4^{\sf const}\,.
\end{equation}
This leads to less energy variation away from thereshold and does not generate the pole near the real axis that arises for the threshold-motivated parameterization. Note that our two scattering parameters have different energy dimensions with $m^9 a_4^{\sf thr}$ and $a_4^{\sf const}$ giving dimensionless quantities.

\begin{figure}
\includegraphics[width=\textwidth]{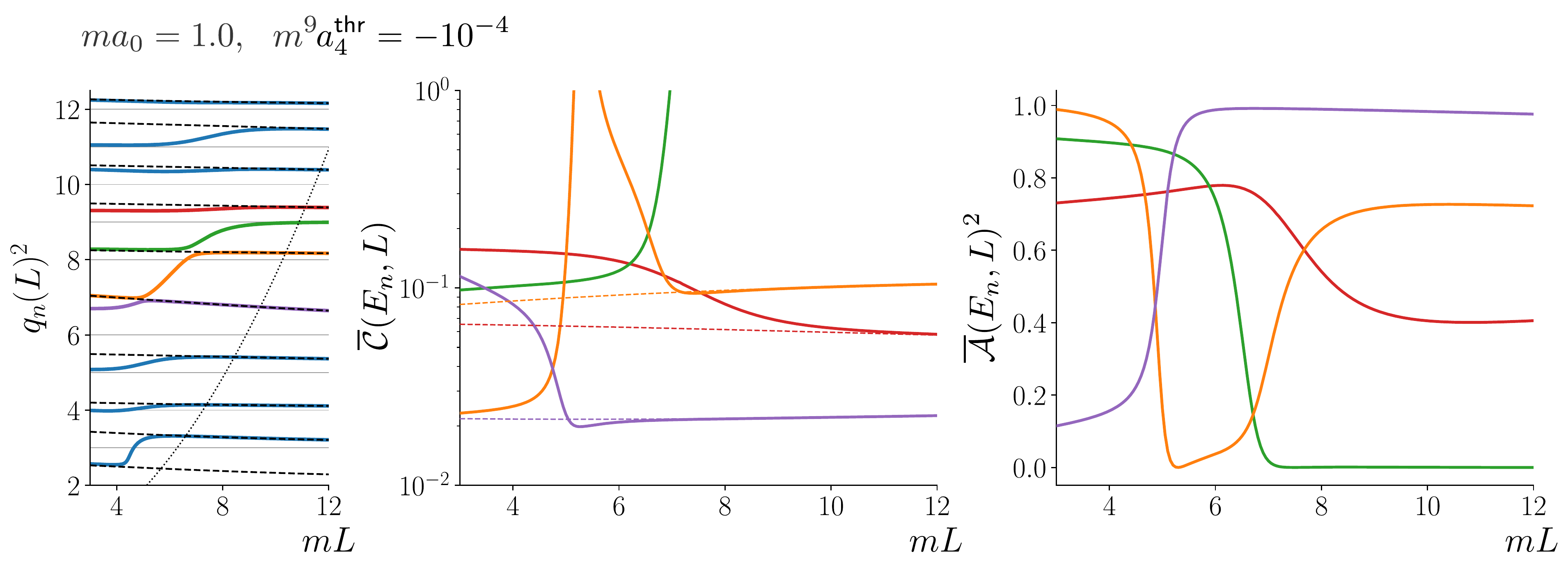}
\caption{Energies and matrix elements exhibiting effects that do not arise in the perturbative expansions for the $\mathcal K^{(4)}_{\sf thr}$ parametrization, defined in eq.~\eqref{eq:g_wave_thr_param}. The left panel shows the finite-volume energies for the full quantization condition plotted vs.~$m L$, using $q_n(L)^2$ defined in eq.~\eqref{eq:q_squared} to express the energies. The middle panel shows the Lellouch- L{\"u}scher factor with non-perturbative $G$-wave effects, summarized in eqs.~\eqref{eq:Cdef} and \eqref{eq:rescaledC}. Finally, the right panel shows the metric of $G$-wave conamination defined in eq.~\eqref{eq:Abar_def}, measuring the fraction of $S$-wave within the finite-volume state.
\label{fig:levels_matelems_NP}}
\end{figure}

Beginning with the threshold-motivated form, in the left panel figure~\ref{fig:levels_matelems_NP} we plot the solutions for the finite-volume energies with the $S$-wave and $G$-wave amplitudes parametrized according to eqs.~\eqref{eq:K0scatteringlengthonly} and \eqref{eq:g_wave_thr_param}, respectively. We set the relevant scattering parameters as $m a_0 = 1.0$ and $m^9 a^{\sf thr}_4 = -10^{-4}$ and solve for the finite-volume energies without making use of any expansion. The order of magnitude for $m^9 a^{\sf thr}_4$ is motivated by the value predicted in chiral perturbation theory for maximal isospin $\pi \pi$ scattering near the kaon mass. As the levels are dense in the region of interest, we find it more instructive to plot
\begin{equation}
\label{eq:q_squared}
q_n(L)^2 = \frac{L^2}{4 \pi^2} \big ( E_n(L)^2/4 - m^2 \big ) \,,
\end{equation}
on the vertical axis. The vicinity of the accidentally degenerate state is highlighted with the three curves that are shown in different colours. We find an additional state relative to the $\ell_{\sf max}=0$ energies (black dashed curves) and that, as $mL$ increases, the energies asymptote to the solutions of the $\ell_{\sf max}=0$ result with the extra solution approaching the non-interacting level at $q_n(L)^2 = 9$. As a given energy level is traced from small to large $mL$ values, it transitions from asymptoting a given $\ell_{\sf max}=0$ solution to asymptoting an adjacent solution.

This complicated spectrum leads to significant variation in the Lellouch-L{\"u}scher factors, as we show in the middle and right panels of figure~\ref{fig:levels_matelems_NP}, focusing on the three states highlighted in different colors in the left panel. For the middle panel, we plot a rescaled version of the quantity $\mathcal C \big (E_{n}(L), L \big )$ appearing in eq.~\eqref{eq:Cdef}:
\begin{equation}
\label{eq:rescaledC}
\overline {\mathcal C} \big (E_{n}(L), L \big ) \equiv \frac{\mathcal C \big (E_{n}(L), L \big ) }{2 E_n(L)^2 L^3} \,.
\end{equation}
This rescaling is motivated by the fact that $\overline{\mathcal C}$ has a straightforward infinite-volume limit as discussed in refs.~\cite{\InfVolLim} and reviewed appendix~\ref{app:large_volume}, see in particular eq.~\eqref{eq:infinite_vol_Cbar}.

We emphasize that no expansion is used in this result; the full $\ell_{\sf max}=4$ result is compared to the result of truncating all expressions at $\ell_{\sf max}=0$, show as the dashed curves. One finds that the $\ell_{\sf max}=0$ and $\ell_{\sf max}=4$ results match for low and high values of $L$, but that, as with the energies, a given factor transitions between two neighbouring $\ell_{\sf max}=0$ counterparts as the volume is varied. The conversion factors exhibit rapid variation in the transition region and can in fact diverge for particular values of $mL$.

To make sense of this behaviour, we return to the relation of eq.~\eqref{eq:LLref} and replace it with a slightly more general form, also derived in ref.~\cite{Briceno:2014uqa}:
\begin{multline}
\Big \vert \sum_{\ell}
\mathcal A_{\ell }(E_{\pi \pi}, L)
\langle E_{\pi \pi}, \pi \pi, \text{out}, \ell \vert \, \mathcal O(0) \, \vert \Phi \rangle
\Big \vert \bigg \vert_{E_{\pi \pi} = E_n(L)} \\
=
\sqrt{2 m_K} L^3 \, \big \vert \langle E_n, L , A_1^+ \vert \mathcal \, O(0) \, \vert \Phi, L \rangle \big \vert \,,
\end{multline}
where we have replaced the weak hamiltonian and kaon ($\mathcal H_W(0) \vert K \rangle$) with a generic operator and single-particle state: $\mathcal O(0) \vert \Phi \rangle$. This is useful to understand the conversion factors more generally, since one can now include combinations of $\mathcal O(0)$ and $\vert \Phi \rangle$ that overlap higher-angular momentum components.

In this case the simple proportionality factor, $\mathcal C \big (E_{n}(L), L \big ) $,
is replaced by the eigenvector $ \mathcal A_{\ell}$, defined as
\begin{equation}
\mathcal A_{\ell' }(E_{n}(L), L) \mathcal A_{\ell }(E_{n}(L), L)^\dagger = \lim_{E \to E_n(L)} \frac{\text{adj} \big [ M^{[A_1^+]}(E, L) \big ]_{\ell' \ell} }{p^{\ell } \cos \delta_\ell (E) \, p^{\ell'} \!\cos \delta_{\ell'} (E)} \,
\bigg ( \frac{\partial \det \! \big [ M^{[A_1^+]}(E, L) \big ] }{\partial E} \bigg)^{-1} \,.
\end{equation}
This makes use of the fact that the ratio of the adjugate to the derivative of the determinant becomes a rank-one matrix when evaluated at a finite-volume energy. There is an inherent phase ambiguity in this definition of $\mathcal A_{\ell}$, and in this work we take the entries to be real and positive. Note that
\begin{equation}
\mathcal A_0(E_n(L), L) = \frac{1}{\sqrt{\mathcal C(E_n(L), L)}} \,,
\end{equation}
whereas the values for $\ell > 0$ are new to this section. In the right panel of figure~\ref{fig:levels_matelems_NP} we plot
the combination
\begin{equation}
\label{eq:Abar_def}
\overline {\mathcal A}(E_n, L)^2 = \frac{\vert \mathcal A_0(E_n(L), L) \vert^2}{\vert \mathcal A_0(E_n(L), L) \vert^2 + \vert \mathcal A_4(E_n(L), L) \vert^2} \,.
\end{equation}
By construction this quantity is in the range $0 < \overline {\mathcal A}(E_n, L)^2 < 1$. This gives a metric for the amount of $G$-wave contamination in a given finite-volume state. As can be seen by comparing the orange curve in the middle and right panels of figure~\ref{fig:levels_matelems_NP}, the divergence in $\overline {\mathcal C}$ corresponds to a point where the $S$-wave component of the finite-volume state vanishes identically.

\begin{figure}
\includegraphics[width=\textwidth]{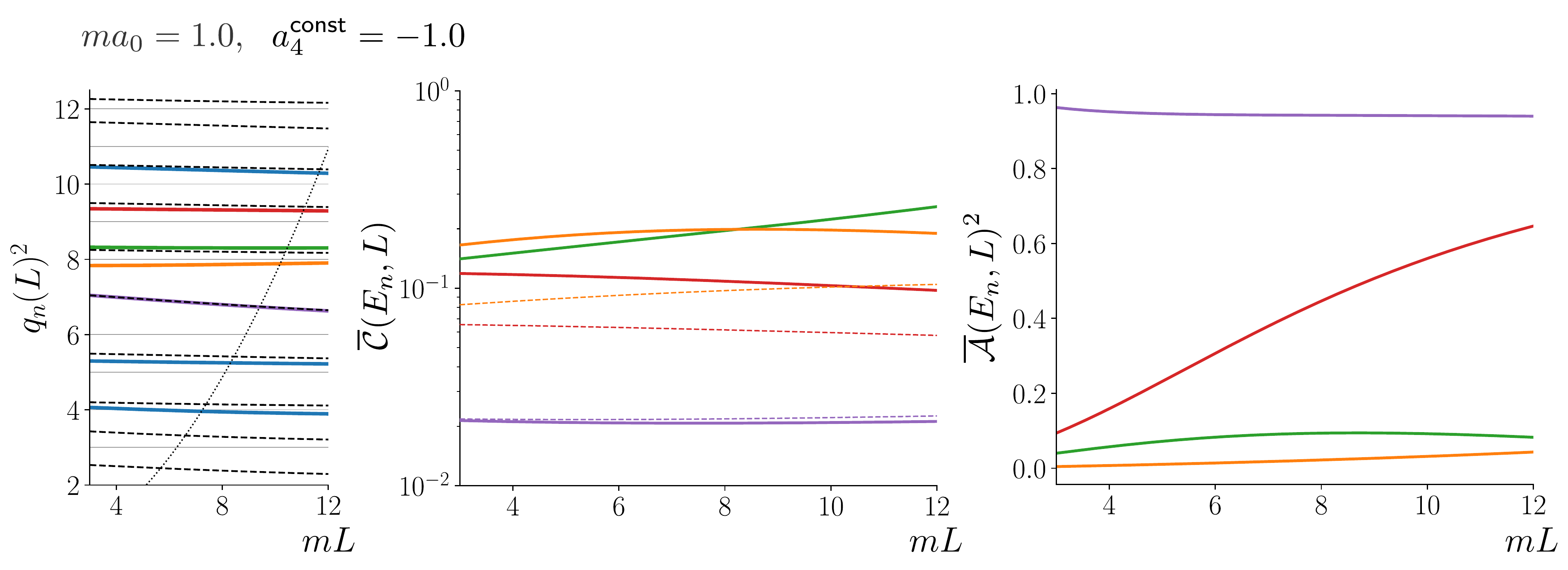}
\caption{Energies and matrix elements exhibiting effects that do not arise in the perturbative expansions for the $\mathcal K^{(4)}_{\sf const}$ parametrization, defined in eq.~\eqref{eq:K4_const_param}. The left, middle and right panels are defined as in figure~\ref{fig:levels_matelems_NP}. \label{fig:levels_matelems_NP_v2}}
\end{figure}

Finally, in figure~\ref{fig:levels_matelems_NP_v2}, we repeat this exercise with our alternative $G$-wave parametrization, given in eq.~\eqref{eq:K4_const_param}. We set the relevant scattering parameters as $m a_0 = 1.0$ and $ a^{\sf const}_4 = -1.0$ as labelled. Note that here no $G$-wave resonance pole is present so the avoided level crossing seen in figure~\ref{fig:levels_matelems_NP} does not arise. We do however continue to observe an extra state in the vicinity of the accidental degeneracy with a significant $G$-wave component.

\section{Results for physical \texorpdfstring{$K \to \pi \pi$}{kaon decays}}
\label{sec:kpipi_realistic}

We now apply the general results of the previous sections to the specific case of the $K \to \pi \pi$ decay amplitude and $\pi \pi \to \pi \pi$ scattering amplitudes. To do so, we require predictions for the elastic scattering phase shifts, or equivalently, the two-to-two K-matrices for isospin zero ($I=0$) and isospin two ($I=2$) two-pion states, each for the lowest three non-zero partial waves, $\ell=0,2,4$. We focus here on physical pion masses in an iso-symmetric theory with degenerate light quarks and without electromagnetism. The iso-symmetric masses are set using the convention of ref.~\cite{RBC:2014ntl}:
\begin{equation}
M_\pi = 135.0 \ \text{MeV} \,, \ \ \ \ \ M_K = 495.7 \ \text{MeV} \,.
\end{equation}

Of the six relevant channels (two isospins $\times$ three angular momenta), all but the scalar-isoscalar ($I=0$, $\ell=0$) are expected to be well approximated by chiral perturbation theory (ChPT). This is because the $I=2$ channel does not contain resonances and, for all four $\ell\neq 0$ channels, the lowest lying resonances appear at centre-of-mass energies much higher than $M_K$. For the case of $I=0$, $\ell=0$, which is the $\sigma$-resonance channel of $\pi \pi$ scattering, we use the phenomenological parametrization of ref.~\cite{Pelaez:2019eqa}.
The parametrization used for the $\sigma$ channel is reviewed in appendix \ref{app:sigma} and the ChPT used for all other channels is summarized in appendix \ref{app:chpt}.

To predict the quantity $\Delta(E, \boldsymbol P, L)$, that gives the $\ell=2$ or $\ell=4$ correction to the Lellouch-L{\"u}scher conversion factor, we first require a lattice set-up for which the $K \to \pi \pi$ decay can be extracted. As is explained in the seminal work of ref.~\cite{\LL}, this requires tuning the finite volume, $M_\pi L$, such that one of the finite-volume energies satisfies $E_n^{\sf cm}(L) = M_K$. Here we expect the effect of the higher-angular momentum to be small in the sense that eq.~\eqref{eq:DeltaEexpanded} will be a useful approximation and, as explained in the text above eq.~\eqref{eq:K0scatteringlengthonly}, this means we can determine the $M_\pi L$ for which $E_n^{\sf cm}(L) = M_K$ using only the $\ell_{\sf max}=0$ quantization condition. To this end, we require the $S$-wave phase shifts evaluated at the kaon mass. Using the parametrization of ref.~\cite{Pelaez:2019eqa} for $I=0$ and the ChPT determination for $I=2$, the $\ell=0$ scattering phase shifts are given by:
\begin{equation}
\begin{split}
I &= 0 : \qquad \delta_{\ell=0}(E_{\sf{cm}}) = 39.13^\circ + 0.18^\circ \bigg ( \frac{E_{\sf{cm}}-M_K}{\text{MeV}} \bigg ) + \mathcal O \left(\left( E_{\sf cm} - M_K\right)^2\right) \,, \\
I &= 2 : \qquad \delta_{\ell=0}(E_{\sf{cm}}) = -11.61^\circ - 0.063^\circ \bigg ( \frac{E_{\sf{cm}}-M_K}{\text{MeV}} \bigg ) + \mathcal O \left(\left( E_{\sf cm} - M_K\right)^2\right) \,.
\end{split}
\end{equation}
Using these values, we determine the volumes required to align a finite-volume energy level with the kaon mass for various choices of total moment $\boldsymbol P$ and twist $\boldsymbol \theta$. The values are summarized in the second columns of tables \ref{tab:isoscalar} and \ref{tab:isotensor}. We do not include uncertainty estimates here as the relevant uncertainties will depend on the details of the lattice calculation. As described more below and in the caption of the table, we do estimate derivatives of $\Delta$ with respect to the phase shifts so that the effect of uncertainties can be quantified.

The tables additionally show the respective values for $\Delta(E,\boldsymbol P,L)$ evaluated using the expansions in eqs.~\eqref{eq:DeltaEexpanded} and \eqref{eq:DeltaExpandedNonzeroP}.
The remaining required inputs, all taken from the ChPT results reviewed in appendix~\ref{app:chpt}, are as follows:
\renewcommand*{\arraystretch}{1.8}
\begin{equation}
\begin{split}
I = 0 : & \qquad \delta_{\ell=2}(E_{\sf{cm}} ) = 0.23^{\circ} + 0.0024^{\circ} \bigg ( \frac{E_{\sf{cm}}-M_K}{\text{MeV}} \bigg ) + \mathcal O \left(\left( E_{\sf cm} - M_K\right)^2\right) ,
\\[2pt]
&\qquad\delta_{\ell=4}(E_{\sf{cm}} ) =1.1\times {10^{-3}}^{\circ} + 1.8\times {10^{-5}}^{\circ} \bigg ( \frac{E_{\sf{cm}}-M_K}{\text{MeV}} \bigg ) + \mathcal O \left(\left( E_{\sf cm} - M_K\right)^2\right) \,, \\[2pt]
I = 2 : & \qquad \delta_{\ell=2}(E_{\sf{cm}} ) =-0.036^{\circ}- 6.9\times {10^{-4}}^{\circ} \bigg ( \frac{E_{\sf{cm}}-M_K}{\text{MeV}} \bigg ) + \mathcal O \left(\left( E_{\sf cm} - M_K\right)^2\right) ,
\\[2pt]
&\qquad\delta_{\ell=4}(E_{\sf{cm}} ) =3.8\times {10^{-4}}^{\circ} + 5.9\times {10^{-6}}^{\circ} \bigg ( \frac{E_{\sf{cm}}-M_K}{\text{MeV}} \bigg ) + \mathcal O \left(\left( E_{\sf cm} - M_K\right)^2\right) \,.
\end{split}
\end{equation}
\renewcommand*{\arraystretch}{1.0}
\renewcommand*{\arraystretch}{1.2}

\begin{table}
\begin{center}
$I = 0$ \\[5pt]
\begin{tabular}{l cccc}
\multicolumn{5}{c}{$\boldsymbol P=[000]$, ${\boldsymbol \theta}=[000]$} \\[3pt]
$n$& $M_\pi L$ & $\Delta$ &
$\partial \Delta/\partial \delta_0 \ [({}^\circ)^{-1}] $ &
$\partial \Delta/\partial \delta_4 \ [({}^\circ)^{-1}] $ \\ \hline
1 & $3.60$ & $1.02 \times 10^{-4}$ & $10^{-6}$ & $-10^{-2}$ \\
2 & $5.36$ & $-1.68 \times 10^{-5}$ & $- 10^{-7}$ & $- 10^{-2}$ \\
3 & $6.89$ & $3.68 \times 10^{-5}$ & $10^{-7}$ & $10^{-2}$ \\
4 & $8.02$ & $6.79 \times 10^{-5}$ & $-10^{-7}$ & $10^{-2}$ \\
5 & $8.76$ & $1.26 \times 10^{-6}$ & $-10^{-7}$ & $10^{-3}$ \\
6 & $9.78$ & $-1.67 \times 10^{-5}$ & $10^{-7}$ & $-10^{-2}$
\end{tabular}
\begin{center}
\begin{tabular}{l cccc}
\multicolumn{5}{c}{$\boldsymbol P=[001]$, ${\boldsymbol \theta}=[000]$} \\[3pt]
$n$& $M_\pi L$ & $\Delta$ &
$\partial \Delta/\partial \delta_0 \ [({}^\circ)^{-1}] $ &
$\partial \Delta/\partial \delta_2 \ [({}^\circ)^{-1}] $ \\ \hline
1 & $4.02$ & $5.65 \times 10^{-4}$ & $10^{-6}$ & $- 10^{-3}$ \\
2 & $5.71$ & $3.68 \times 10^{-3}$ & $- 10^{-4}$ & $ 10^{-2}$ \\
3 & $5.96$ & $-0.101$ & $- 10^{-3}$ & $-10^{-1}$ \\
4 & $7.03$ & $2.74 \times 10^{-3}$ & $10^{-5}$ & $ 10^{-2}$ \\
5 & $8.08$ & $-1.17 \times 10^{-3}$ & $- 10^{-5}$ & $- 10^{-2}$ \\
6 & $8.37$ & $-0.239$ & $- 10^{-3}$ & $-10^{0}$
\end{tabular}
\end{center}
\begin{center}
\begin{tabular}{l cccc}
\multicolumn{5}{c}{$\boldsymbol P=[000]$, ${\boldsymbol \theta}=[\frac{1}{2}\frac{1}{2}\frac{1}{2}]$} \\[3pt]
$n$& $M_\pi L$ & $\Delta$ &
$\partial \Delta/\partial \delta_0 \ [({}^\circ)^{-1}] $ &
$\partial \Delta/\partial \delta_4 \ [({}^\circ)^{-1}] $ \\ \hline
1 & $2.78$ & $-9.38 \times 10^{-5}$ & $-10^{-5}$ & $-10^{0}$ \\
2 & $6.12$ & $3.42 \times 10^{-6}$ & $-10^{-8}$ & $-10^{-3}$ \\
3 & $8.53$ & $8.02 \times 10^{-6}$ & $10^{-8}$ & $10^{-3}$ \\
4 & $10.2$ & $2.61 \times 10^{-6}$ & $10^{-8}$ & $-10^{-3}$ \\
5 & $11.7$ & $-2.78 \times 10^{-6}$ & $-10^{-8}$ & $-10^{-3}$
\end{tabular}
\end{center}
\end{center}
\caption{\label{tab:isoscalar}Table of $M_\pi L$ values for which $E^{\sf cm}_n(L) = M_K$, in the $I=0$ $\pi \pi $ system, with the excited state index $n$, the total momentum $\boldsymbol P$, and the boundary conditions $\boldsymbol{\theta}$ as specified. The second column in each table gives $\Delta$, which is the relative correction relating the $S$-wave only and the full Lellouch-L{\"u}scher factors, as defined in eq.~\eqref{eq:first_delta_def}. The final two columns give derivatives of this correction with respect to the phase shifts. Here we only give the order of magnitude since the intent is to give a sense of how uncertainty would propagate in the correction. As is consistent with the expansion used in eqs.~\eqref{eq:DeltaEexpanded} and \eqref{eq:DeltaExpandedNonzeroP}, the box sizes for which $E^{\sf cm}_n(L) = M_K$ are determined using only the $\ell=0$ partial wave.}
\end{table}

\begin{table}
\begin{center}
$I = 2$ \\[5pt]
\begin{tabular}{l cccc}
\multicolumn{5}{c}{$\boldsymbol P=[000]$, ${\boldsymbol \theta}=[000]$} \\[3pt]
$n$& $M_\pi L$ & $\Delta$ &
$\partial \Delta/\partial \delta_0 \ [({}^\circ)^{-1}] $ &
$\partial \Delta/\partial \delta_4 \ [({}^\circ)^{-1}] $ \\ \hline
1 & $4.20$ & $2.00 \times 10^{-5}$ & $10^{-7}$ & $-10^{-2}$ \\
2 & $5.90$ & $-2.98 \times 10^{-6}$ & $-10^{-7}$ & $-10^{-2}$ \\
3 & $7.13$ & $4.92 \times 10^{-6}$ & $10^{-7}$ & $10^{-3}$ \\
4 & $8.19$ & $-1.60 \times 10^{-6}$ & $10^{-6}$ & $-10^{-2}$ \\
5 & $9.22$ & $5.04 \times 10^{-6}$ & $-10^{-7}$ & $10^{-2}$ \\
6 & $10.1$ & $-2.89 \times 10^{-6}$ & $-10^{-7}$ & $-10^{-2}$
\end{tabular}
\end{center}
\begin{center}
\begin{tabular}{l cccc}
\multicolumn{5}{c}{$\boldsymbol P=[001]$, ${\boldsymbol \theta}=[000]$} \\[3pt]
$n$& $M_\pi L$ & $\Delta$ &
$\partial \Delta/\partial \delta_0 \ [({}^\circ)^{-1}]$ &
$\partial \Delta/\partial \delta_2 \ [({}^\circ)^{-1}] $ \\ \hline
1 & $4.63$ & $-1.40 \times 10^{-4}$ & $-10^{-6}$ & $-10^{-4}$ \\
2 & $5.89$ & $-1.79 \times 10^{-2}$ & $10^{-4}$ & $10^{-1}$ \\
3 & $6.17$ & $2.25 \times 10^{-4}$ & $10^{-5}$ & $-10^{-2}$ \\
4 & $7.27$ & $-4.93 \times 10^{-6}$ & $-10^{-5}$ & $-10^{-2}$ \\
5 & $8.35$ & $-1.45 \times 10^{-2}$ & $10^{-3}$ & $10^{-1}$ \\
6 & $8.46$ & $7.48 \times 10^{-4}$ & $10^{-4}$ & $-10^{-2}$
\end{tabular}
\end{center}
\begin{center}
\begin{tabular}{l cccc}
\multicolumn{5}{c}{$\boldsymbol P=[000]$,
${\boldsymbol \theta}=[\frac{1}{2}\frac{1}{2}\frac{1}{2}]$} \\[3pt]
$n$& $M_\pi L$ & $\Delta$ &
$\partial \Delta/\partial \delta_0 \ [({}^\circ)^{-1}]$ &
$\partial \Delta/\partial \delta_4 \ [({}^\circ)^{-1}] $ \\ \hline
1 & $3.76$ & $-1.25 \times 10^{-6}$ & $-10^{-7}$ & $-10^{-2}$ \\
2 & $6.95$ & $1.58 \times 10^{-6}$ & $-10^{-8}$ & $10^{-3}$ \\
3 & $9.00$ & $1.57 \times 10^{-6}$ & $10^{-8}$ & $10^{-4}$ \\
4 & $10.7$ & $3.78 \times 10^{-7}$ & $10^{-8}$ & $-10^{-3}$

\end{tabular}
\end{center}
\caption{\label{tab:isotensor}As in table~\ref{tab:isoscalar} but for the $I=2$ $\pi \pi $ system.}
\end{table}

We find that the contamination from the $G$-wave in the extraction of the $K \to \pi \pi$ decay amplitude is at the sub-per-mille level, making it a negligible source of systematic uncertainty for present day calculations. This is true, in particular, for the calculations performed to date by the RBC/UKQCD collaboration. There are, however, a few specific cases where the effect can reach the few-percent level and potentially become relevant. The most extreme example is the case of periodic boundary conditions in which the second excited state is tuned to match the kaon mass with $\boldsymbol P = [001]$, resulting in a volume of about $M_\pi L \sim 6$. Here we find an order $10 \%$ correction arising from the $D$-wave for the $I=0$ case. A final word of warning regarding the $\Delta$ values tabulated here is that these are very sensitive to the precise parameter values chosen. Thus, in a calculation approaching the precision where these effects are relevant, one would potentially require a large systematic uncertainty on the correction factor.

\section{Conclusion}
\label{sec:conc}

In this work we have studied the phenomenology of the formulae, derived in refs.~\cite{Lellouch:2000pv,Briceno:2014uqa}, for extracting $1 \overset{\mathcal J}{\to} 2$ transition amplitudes from finite-volume matrix elements. In section \ref{sec:fvformalism} we review the standard formalism and in section~\ref{sec:exp_num} we derive expansions to correct the standard ($S$-wave-only) Lellouch-L{\"u}scher factors at leading order in the $2 \to 2$ K-matrix of the next contaminating partial wave. We also give numerical results to estimate the size of the contamination over a range of possible interaction strengths, in particular in figures~\ref{fig:P000_periodic}, \ref{fig:P000_antiperiodic}, and \ref{fig:P001_periodic}.

In section~\ref{sec:exp_num}, we then turn to the investigation of finite-volume phenomena, predicted by the general formalism in a regime where perturbative expansions are no longer valid. We illustrate two specific effects here: first, the splitting of the accidentally degenerate 8${}^\text{th}$ excited state, and second, the erratic behaviour of the finite-volume conversion factor for the case of a resonance in the higher partial wave.

Finally, in section~\ref{sec:kpipi_realistic} we tabulate numerical results for the relative size of higher-partial wave in the case of a physical $K \to \pi \pi$ lattice calculation. These estimates may be useful for future $K \to \pi \pi$ lattice calculations, targeting increased precision in this phenomenologically important weak decay.

\acknowledgments
We warmly thank Mattia Bruno, Dorota Grabowska, Andrew Jackura, Fernando Romero-L{\'o}pez, Akaki Rusetsky, and Steve Sharpe for useful discussions. We especially thank Arkaitz Rodas for providing guidance on the use of the parametrization of ref.~\cite{Pelaez:2019eqa}. MTH is supported by UKRI Future Leaders Fellowship MR/T019956/1, and both MTH and TP are supported in part by UK STFC grant ST/P000630/1.

\appendix

\section{Details of the finite-volume functions}

\label{app:Ffuncs}
The $\widetilde{F}$ matrices that enter the quantization condition, eq.~\eqref{eq:Mdef}, are given by \cite{Kim:2005gf}:

\begin{equation}
\widetilde F_{\ell m, \ell' m'}(E, \boldsymbol P, L) = \frac12 \bigg [ \frac{1}{L^3} \sum_{\boldsymbol k \in \frac{2 \pi}{L} \mathbb Z^3} - \, \text{p.v.} \! \int \! \! \frac{d^3 \boldsymbol k}{(2 \pi)^3} \bigg ] \frac{\mathcal Y_{\ell m}(\boldsymbol k_{\sf{cm}}) \mathcal Y^*_{\ell' m'}(\boldsymbol k_{\sf{cm}}) }{2 \omega_{\boldsymbol k} 2 \omega_{\boldsymbol P - \boldsymbol k} (E - \omega_{\boldsymbol k} - \omega_{\boldsymbol P - \boldsymbol k})} \,,
\end{equation}
where
\begin{align}
\mathcal Y_{\ell m}(\boldsymbol k_{\sf{cm}}) & = \sqrt{4 \pi} \vert \boldsymbol k_{\sf{cm}} \vert^{\ell} Y_{\ell m}(\hat {\boldsymbol k}_{\sf{cm}}) \,, \\
\omega_{\boldsymbol k} & = \sqrt{\boldsymbol k^2 + m^2} \\
\omega_{\boldsymbol P - \boldsymbol k} & = \sqrt{(\boldsymbol P - \boldsymbol k)^2 + m^2} \,, \\
\boldsymbol k_{\sf{cm}} & = \boldsymbol \beta \gamma \omega_{\boldsymbol k} + \gamma \Big ( \hat {\boldsymbol \beta} ( \boldsymbol k \cdot \hat {\boldsymbol \beta} ) \Big ) + \Big ( \boldsymbol k - \hat {\boldsymbol \beta} ( \boldsymbol k \cdot \hat {\boldsymbol \beta} ) \Big ) \bigg \vert_{\gamma = E/E_{\sf{cm}}, \ \boldsymbol \beta = - \boldsymbol P/E}\,.
\end{align}
Note that up to this stage the expression is general, that is it holds for moving frames and twisted boundary conditions.

The relevant entries are nicely summarized in ref.~\cite{Luu:2011ep}. In the case of no twisting and no overall momentum the relevant contributions are:
\begin{align}
\overline F_{00}(E,L) & = -\frac{1}{16 \pi E } \frac{2 }{\sqrt{\pi} L}\mathcal{Z}_{0,0} ( q^2 ) \,, \\
\overline F_{40}(E,L) & = \overline F_{04}(E,L) = - \frac{1}{16 \pi E} \bigg ( \frac{2 \pi}{L} \bigg )^{\! \!4} \, \frac{2 \sqrt{3} }{\sqrt{7} } \frac{2}{\sqrt{\pi} L} \mathcal{Z}_{4,0} ( q^2 ) \,, \\
\begin{split}
\overline F_{44}(E,L) & = - \frac{p^8}{16 \pi E } \frac{2 }{\sqrt{\pi} L}
\bigg [
\mathcal{Z}_{0,0}( {q}^2)
+
\frac{108}{143 q^4} \mathcal{Z}_{4,0} ( {q}^2 )
\\ & \hspace{115pt} +
\frac{80}{11 \sqrt{13} q^6 } \mathcal{Z}_{6,0} ( {q}^2 ) +
\frac{560 }{143
\sqrt{17} q^8 } \mathcal{Z}_{8,0} ({q}^2 )
\bigg ] \,,
\end{split}
\end{align}
where $q = p L/(2 \pi)$ and
\begin{equation}
\mathcal{Z}_{\ell,m} ( {q}^2 ) \equiv \mathcal{Z}_{\ell,m} (1 , {q}^2 ) \,, \qquad \qquad \mathcal{Z}_{\ell,m} (s , {q}^2 ) \equiv \sum_{\boldsymbol n} \frac{\vert \boldsymbol n \vert^{\ell} Y_{\ell m}(\hat {\boldsymbol n})}{[\boldsymbol n^2 - q^2]^s} \,,
\end{equation}
with analytic continuation from $s > 3/2$ providing a regulator for the $\ell = m = 0$ function. Note that throughout this work we use $k$ a summed momentum, $p$ as a physical momentum magnitude (corresponding to $E$), and $q$ as a dimensionless version of $p$. A convenient numerical expression for $\mathcal Z_{\ell, m}$ is given by:
\begin{align}
\mathcal{Z}_{\ell,m} (1 , {q}^2 ) & = \lim_{\alpha \to 0^+} \bigg ( \sum_{\boldsymbol n} \frac{\vert \boldsymbol n \vert^{\ell} Y_{\ell m}(\hat {\boldsymbol n})}{\boldsymbol n^2 - q^2 } e^{\alpha (q^2 - \boldsymbol n^2)} - \delta_{\ell 0} \delta_{m0} \, \mathcal I(\alpha, q^2) \bigg )
\,,
\\
\mathcal I(\alpha, q^2) & \equiv \sqrt{4 \pi}
\bigg [ \frac{\sqrt{\pi } e^{\alpha q^2}}{2 \sqrt{\alpha}}-\frac{1}{2} \pi \sqrt{q^2} \, \text{erfi} (\sqrt{\alpha q^2} ) \bigg ] \,.
\end{align}

For the case of ${\boldsymbol \theta} = (\frac12, \frac12, \frac12)$ the relevant components of $\overline F$ are the same as above, with the difference that the sum in $\mathcal{Z}$ is over the twisted momenta instead. This is accounted for by making the replacement:
\begin{equation}
\mathcal Z_{\ell,m}(s, q_{\sf{cm}}^2)\to \mathcal Z_{\ell, m}^{{\boldsymbol \theta}}(s, q_{\sf{cm}}^2) = \sum_{\boldsymbol r \in \mathcal P_{{\boldsymbol \theta}}} \frac{\vert \boldsymbol r \vert^{\ell} Y_{\ell m}(\hat {\boldsymbol r}) }{[\boldsymbol r^2 - q^2_{\sf{cm}}]^s}\,,
\end{equation}
where
\begin{equation}
\mathcal P_{{\boldsymbol \theta}} = \Big \{\boldsymbol r \in \mathbb R^3 \, \Big \vert \, \boldsymbol r = \boldsymbol n + {\boldsymbol \theta}, \, \boldsymbol n \in \mathbb Z^3 \Big \} \,.
\end{equation}

Note that this simple replacement is only valid for our particular choice of twist, if instead the twist was, for example, ${\boldsymbol \theta}=(0,0,1)$ then $\ell=2$ components would also contribute.

Finally, focusing on the case $\boldsymbol d = [001]$ (recall $\boldsymbol P = (2 \pi/L)\boldsymbol d$), this gives
\begin{align}
\overline F_{00}(E,\boldsymbol P,L) & = - \frac{1}{16 \pi E_{\sf{cm}} } \frac{2 }{\sqrt{\pi} L \gamma}\mathcal{Z}^{\boldsymbol d}_{0,0} ( q_{\sf{cm}}^2, L ) \,, \\
\overline F_{20}(E,\boldsymbol P,L) & = \overline F_{02}(E,\boldsymbol P,L) = - \frac{1}{16 \pi E_{\sf{cm}} } \bigg ( \frac{2 \pi}{L} \bigg )^2 \frac{2 }{\sqrt{\pi} L \gamma}\mathcal{Z}^{\boldsymbol d}_{2,0} ( q_{\sf{cm}}^2, L ) \,,
\\
\begin{split}
\overline F_{22}(E,\boldsymbol P,L) & = - \frac{p_{\sf{cm}}^4}{16 \pi E_{\sf{cm}} } \frac{2 }{\sqrt{\pi} L \gamma}
\bigg [
\mathcal{Z}^{\boldsymbol d}_{0,0} ( q_{\sf{cm}}^2,L )
+
\frac{2 \sqrt{5}}{7 q_{\sf{cm}}^2}
\mathcal{Z}^{\boldsymbol d}_{2,0} ( q_{\sf{cm}}^2,L )
+
\frac{6}{7 q_{\sf{cm}}^4 } \mathcal{Z}^{\boldsymbol d}_{4,0} ( q_{\sf{cm}}^2,L )
\bigg ] \,,
\end{split}
\end{align}
where $\gamma = E/E_{\sf{cm}}$ is the standard Lorentz boost vector associated with the total energy and momentum $(E, \boldsymbol P)$.

Here $ \mathcal{Z}^{\boldsymbol d}_{\ell,m} $ is a moving frame version of the zeta-function defined, for any $\boldsymbol d$, as
\begin{align}
\mathcal{Z}^{\boldsymbol d}_{\ell,m} (s, q_{\sf{cm}}^2 , L) & = \sum_{\boldsymbol r \in \mathcal P_{\boldsymbol d}} \frac{\vert \boldsymbol r \vert^{\ell} Y_{\ell m}(\hat {\boldsymbol r}) }{[\boldsymbol r^2 - q^2_{\sf{cm}}]^s} \,,
\\
& = \lim_{\alpha\to 0^+}\bigg(\sum_{\boldsymbol r \in \mathcal P_{\boldsymbol d}} \frac{\vert \boldsymbol r \vert^{\ell} Y_{\ell m}(\hat {\boldsymbol r}) }{\boldsymbol r^2 - q^2_{\sf{cm}}}e^{\alpha (q_{\sf{cm}}^2-r^2)} - \gamma \, \delta_{\ell 0} \delta_{m0} \,
\mathcal I(\alpha, q^2_{\sf{cm}}) \bigg) \,,
\end{align}
where the set $\mathcal P_{\boldsymbol d}$ is given by
\begin{equation}
\mathcal P_{\boldsymbol d} = \Big \{\boldsymbol r \in \mathbb R^3 \, \Big \vert \, \boldsymbol r = \hat {\boldsymbol \gamma}^{-1} \big [ \boldsymbol n - {\boldsymbol d}/2 \big ] , \, \boldsymbol n \in \mathbb Z^3 \Big \} \,,
\end{equation}
where $\hat {\boldsymbol \gamma}^{-1}$ acts by re-scaling the component parallel to $\boldsymbol d$ with the inverse boost factor:
\begin{equation}
\hat {\boldsymbol \gamma}^{-1} \boldsymbol x = \frac{1}{\gamma} \boldsymbol d \frac{\boldsymbol x \cdot \boldsymbol d}{\boldsymbol d^2} + \Big ( \boldsymbol x - \boldsymbol d \frac{\boldsymbol x \cdot \boldsymbol d}{\boldsymbol d^2} \Big ) \,,
\end{equation}
for a generic three-vector $\boldsymbol x$. For the special case of $\boldsymbol d = [001]$ this corresponds to to
\begin{equation}
\hat {\boldsymbol \gamma}^{-1} =\text{diag}\left(1,1,\frac{1}{\gamma}\right) \,.
\end{equation}

\section{Large-volume limit}
\label{app:large_volume}

\begin{figure}
\centering
\includegraphics[width=0.5\textwidth]{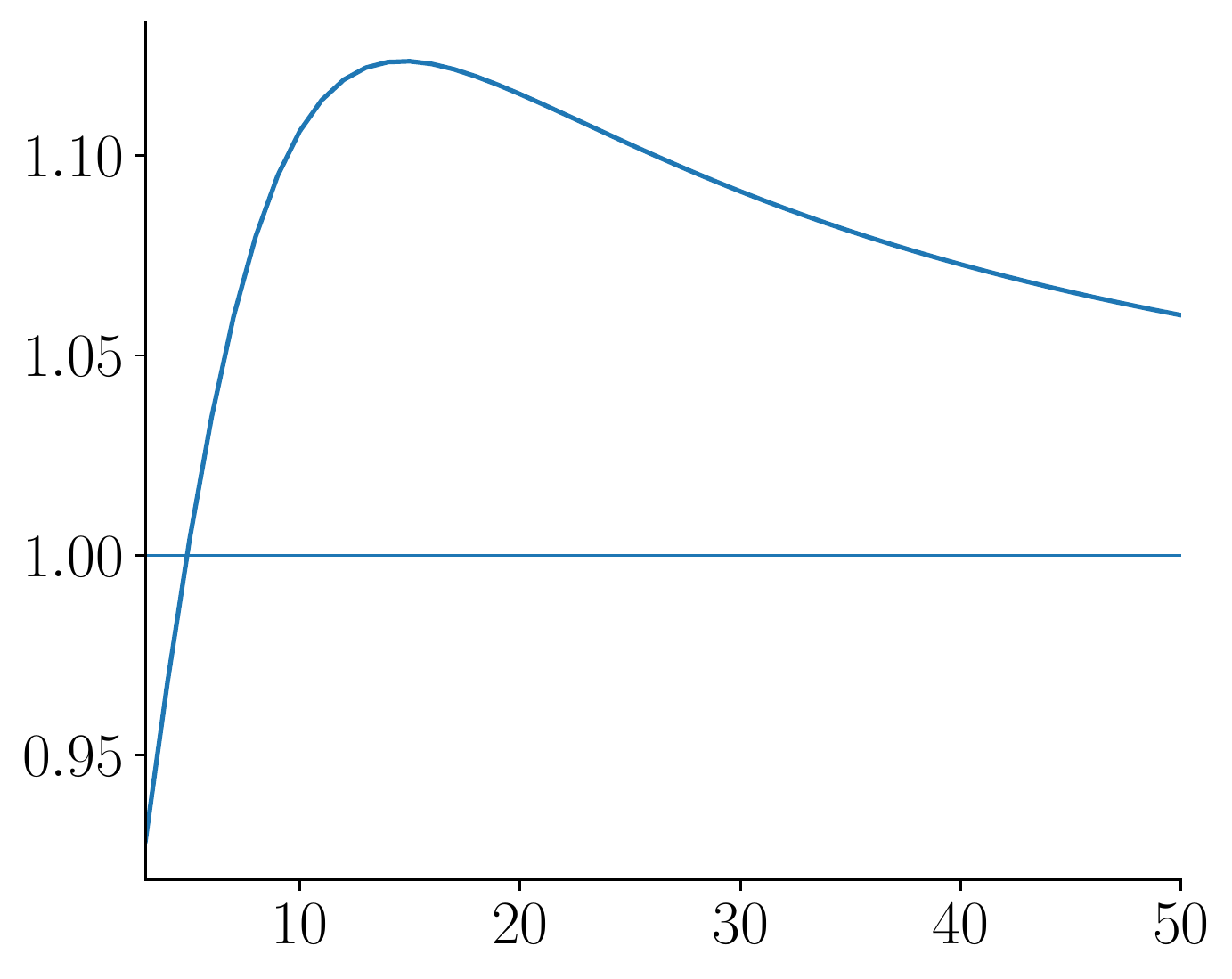}
\caption{A plot of $1/[12 \overline {\mathcal C} \big (E_{n=2}(L), L \big )]$ showing the approach to the infinite-volume limit.\label{fig:infinite_volume_lim}}
\end{figure}

In this appendix we evaluate the large-volume limit for the finite-volume conversion factor, \linebreak $\mathcal C(E_n(L), L)$, defined in eq.~\eqref{eq:Cdef} of the main text. We restrict attention to the case of zero spatial momentum: $\boldsymbol P=[000]$.

In the large-volume limit, it is useful to write the interacting energy as
\begin{align}
\label{eq:en_large_vol_v1}
E_n(L) = 2\omega_n(L) + \epsilon_n(L) \,,
\end{align}
where we have introduced
\begin{equation}
\label{eq:omega_n_def}
\omega_n(L) = \sqrt{m^2+ \boldsymbol n_n^2 (2\pi/L)^2 } \,,
\end{equation}
to parametrize the non-interacting energy and $\epsilon_n(L)$ as the deviation between interacting and non-interacting energies. A key result that we will make use of below is that $\epsilon_n(L) = \mathcal O(1/L^3)$ \cite{Grabowska_2022}.
In eq.~\eqref{eq:omega_n_def}, $\boldsymbol n^2_n$ is the squared magnitude of a three-vector of integers, corresponding to the $n^\text{th}$ non-interacting energy level. The vector $\boldsymbol n_n$ is not uniquely defined but here we only require the magnitude, so the ambiguity is irrelevant. At this stage, we also assume the state under consideration is not accidentally degenerate. We will return to the case of accidental degeneracy at the end of the appendix.

In turns out that the infinite-volume limit of $\mathcal C(E_n(L), L)$ itself in fact diverges as $L^3$, so we will instead evaluate the limit on the rescaled quantity $\overline{\mathcal C}(E_n(L), L)$, defined in eq.~\eqref{eq:rescaledC} of the main text. We repeat the definition here, for convenience
\begin{equation}
\overline {\mathcal C} \big (E_{n}(L), L \big ) = \frac{\mathcal C \big (E_{n}(L), L \big ) }{2 E_n(L)^2 L^3} \,.
\end{equation}
Combining this with eq.~\eqref{eq:Cdef} we write
\begin{align}
\lim_{L \to \infty} \overline {\mathcal C} \big (E_{n}(L), L \big ) & = \lim_{L \to \infty} \frac{\cos^2 \delta_0 (E_n(L))}{2 E_n(L)^2 L^3} \Big [ \partial_E \overline F_{00}(E,L)^{-1}+ \partial_E \mathcal K^{(0)} (E) \Big ]_{E = E_n(L)}\,, \\[10pt]
& = - \lim_{L \to \infty} \frac{1}{2 E_n(L)^2 L^3} \overline F_{00}(E_n(L),L)^{-2} \partial_E \overline F_{00}(E,L) \Big \vert_{E = E_n(L)} \,,
\label{eq:limit_step_two}
\end{align}
where in the second line we used that $\partial_E \mathcal K^{(0)} (E) \big \vert_{E = E_n(L)}$ remains finite as $L \to \infty$ so that the second term does not contribute and additionally that $\delta(E_n(L)) \to 0$ as $L \to \infty$.

To complete the evaluation, we next study $F_{00}(E,L)$ evaluated at $E_n(L)$, making use of the parametrization given in eq.~\eqref{eq:en_large_vol_v1}
\begin{align}
\overline F_{00}(E_n(L), L) & = \frac12 \bigg [ \frac{1}{L^3} \sum_{\boldsymbol k \in \frac{2\pi}{L} \mathbb{Z}^3} - \, \text{p.v.} \! \int \! \! \frac{d^3 \boldsymbol k}{(2 \pi)^3} \bigg ] \frac{1}{(2 \omega_{\boldsymbol k})^2 (E_n(L)-2\omega_{\boldsymbol{k}})}\\
&= \frac12 \bigg [ \frac{1}{L^3} \sum_{\boldsymbol k \in \frac{2\pi}{L} \mathbb{Z}^3} - \, \text{p.v.} \! \int \! \! \frac{d^3 \boldsymbol k}{(2 \pi)^3} \bigg ] \frac{1}{(2 \omega_{\boldsymbol k})^2 (2\omega_n-2\omega_{\boldsymbol{k}}+\epsilon_n(L))}\\
\begin{split}
&=\frac{1}{L^3} \sum_{\vert \boldsymbol k \vert = (2 \pi/L) \vert \boldsymbol n_n \vert} \frac{1}{8 \omega_n(L)^2 \epsilon_n(L)}
\\[5pt]
& \quad+ \frac12 \bigg [ \frac{1}{L^3} \sum_{\vert \boldsymbol k \vert\neq (2 \pi/L) \vert \boldsymbol n_n \vert} - \, \text{p.v.} \! \int \! \! \frac{d^3 \boldsymbol k}{(2 \pi)^3} \bigg ] \frac{1}{(2 \omega_{\boldsymbol k})^2 (2 \omega_n - 2 \omega_{\boldsymbol k} + \epsilon_n(L) )} \,,
\end{split}
\end{align}
where we have split the quantity according to the $\epsilon_n(L)$ scaling of the terms. The key point is that, while the first line is non-zero in the $L \to \infty$ limit, the second line vanishes. The first line survives exactly due to the $\epsilon_n(L) \sim 1/L^3$ scaling mentioned above, while the second line tends to zero because the sum with the singularity removed approaches the principal value integral and thus is cancelled by the subtracted integral.

This analysis can be repeated for the energy derivative of $\overline{F}_{00}(E, L)$ where the relevant term can be found by differentiating the above with respect to $\epsilon_n(L)$. Combining the surviving contributions into eq.~\eqref{eq:limit_step_two}, we reach
\begin{align}
\begin{split}
\lim_{L \to \infty} \overline {\mathcal C} \big (E_{n}(L), L \big ) & = \lim_{L \to \infty} \frac{1}{2 (2 \omega_n(L) )^2 } \bigg [ \frac{1}{L^3} \sum_{\vert \boldsymbol k \vert = (2 \pi/L) \vert \boldsymbol n_n \vert} \frac{1}{8 \omega_n(L)^2 \epsilon_n(L)} \bigg ]^{-2} \\[5pt]
& \hspace{130pt} \times
\frac{1}{L^3} \sum_{\vert \boldsymbol k \vert = (2 \pi/L) \vert \boldsymbol n_n \vert} \frac{1}{8 \omega_n(L)^2 \epsilon_n(L)^2}\,,
\end{split} \\
& = \bigg [\sum_{\vert \boldsymbol k \vert = (2 \pi/L) \vert \boldsymbol n_n \vert} \bigg]^{-1} \,,
\end{align}
where in the second line we have used that the summand does not depend on the summed coordinate and that all factors cancel except for a single sum as shown. This sum runs over the set of integer vectors with the indicated magnitude and the final result can thus be written as
\begin{equation}
\label{eq:infinite_vol_Cbar}
\lim_{L \to \infty} \overline {\mathcal C} \big (E_{n}(L), L \big ) = \frac{1}{\nu_n} \,,
\end{equation}
where $\nu_n$ is the multiplicity of the $n^{\text{th}}$ non-interacting state, e.g.~$\nu_0 = 1,\nu_1 = 6,\nu_2 = 12, \cdots$.

\section{Parametrization of \texorpdfstring{$\pi \pi \to \pi \pi$}{pion scattering}}

\subsection{Phenomenological description for the scalar iso-scalar sector}
\label{app:sigma}

Due to the $f_0(500)$ resonance in the $I=0$, $\ell=0$ channel, standard chiral perturbation theory is not guaranteed to give a good description of $\pi\pi$ scattering in this channel for two-pion energies near the kaon mass. For this reason we have decided to instead use the phenomenological parametrization provided by ref.~\cite{Pelaez:2019eqa}.

Equation (6) of that paper (together with a few definitions in the text and neighbouring equations) states that the scattering phase shift below the two-kaon threshold (i.e.~for $s = E^2_{\sf{cm}} < (2 M_K)^2$, where $M_K$ is the kaon mass) is well described by the following parametrization:
\begin{itemize}
\item The scattering amplitude---referred to in ref.~\cite{Pelaez:2019eqa} as $t$ instead of $\mathcal M$---is related to the scattering phase shift as and split up as:
\begin{align}
t(s)=\frac{1}{\sigma_\pi(s)}\frac{1}{\cot \delta^0_0(s)-i}=t_{conf}(s)+t_{f_0}(s)+2i\sigma_\pi(s)t_{conf}(s)t_{f_0}(s) \,,
\end{align}
where $\sigma(s)$ is the phase space factor:
\begin{align}
\sigma_i(s)=\frac{2 q_i}{\sqrt{s}}=\sqrt{1-\frac{4m_i^2}{s}}.
\end{align}
Note that as we are only considering the isospin-zero, $S$-wave scattering amplitude we have removed those labels as compared with the original paper.
\item The contribution $t_{conf}$ is given by:
\begin{align}
t_{conf}(s)=\frac{1}{\sigma_\pi(s)}\frac{1}{\Phi(s)-i}
\end{align}
where the function
\begin{align}
\Phi(s)=\frac{\sqrt{s}}{\sqrt{s - 4 M_\pi^2}}\frac{M_\pi^2}{s-z_0^2/2} \left[\frac{z_0^2}{M_\pi \sqrt{s}}+\sum_{n=0}^N B_n\left(\frac{\sqrt{s}-\sqrt{4 M_K^2-s}}{\sqrt{s}+\sqrt{4 M_K^2-s}}\right)^{\!\!\!n} \ \right] \,,
\end{align}
and the parameter values are given below.
\item The final contribution $t_{f_0}$ is given by:
\begin{align}
t_{f_0}(s)=\frac{s G}{M-s-J_\pi(s)s G-J_K(s)m_K^2f(s)} \,,
\end{align}
where
\begin{align}
J_i(s)&=\frac{2}{\pi}+\frac{\sigma_i(s)}{\pi}\log\left(\frac{\sigma_i(s)-1}{\sigma_i(s)+1}\right)\\
f(s)&=\sum_{n=0}^{\tilde{N}} K_n \ x_n \bigg (2\frac{\sqrt{s}-2 M_K}{1.5\text{GeV}-2M_K}-1 \bigg ) \,, \\[5pt]
\begin{split}
M&=\frac{(f^IJ_K^R+f^RJ_K^I)(s^I(J_\pi^I-2\sigma_\pi^R)-s^R(J_\pi^R+2\sigma_\pi^I))+(J_\pi^I-2\sigma_\pi^R)((s^I)^2+(s^R)^2)}{J_\pi^Is^R+J_\pi^Rs^I+2(\sigma_\pi^Is^I-s^R\sigma_\pi^R)}\\&\qquad \qquad-(f^IJ_K^I-f^RJ_K^R) \,,
\end{split}\\
G&=-\frac{f^IJ_K^R+F^RJ_K^I+s^I}{J_\pi^Is^R+J_\pi^Rs^I+2(\sigma_\pi^Is^I-s^R\sigma_\pi^R)} \,,
\end{align}
where $x_i$ is the Chebyshev polynomial of order $i$, the parameters $K_i$ are given below and in $M$ and $G$ all functions of $s$ are evaluated at $s_p$. The superscripts $R,\ I$ signify either the real or imaginary part. We keep all reference to the particle mass as a subscript, as opposed to an argument as in the original paper, and have already made relevant substitutions to shorten the presentation.
\end{itemize}

The authors provide several choices for the parameter values. In this work we use parameter values of Solution I given in the left column of Table 1. We repeat the relevant values here for convenience:
\begin{align}
& N = 5 \,, && z_0=0.137\pm0.028 \, \text{GeV} \,, &&\nonumber \\
&B_0=12.2\pm 0.3, && B_1=-0.9\pm 1.1,&& B_2=15.9\pm 2.7,
\\
& B_3=-5.7\pm 3.1, &&B_4=-22.5\pm 3.7,&& B_5=6.9\pm 4.8 \,\nonumber\\
&\tilde{N}=3 \,, &&\sqrt{s_p}=0.996\pm 7-(0.025\pm8)i \ \text{GeV} \,,&&\nonumber\\
& K_0=5.25\pm0.28,&&K_1=-4.40\pm0.16,&&K_2=0.175\pm0.155,\\
&K_3=-0.28\pm0.06.&&\nonumber
\end{align}
For the parametrization we take $M_\pi=139.57 \,{\rm MeV}$ as in ref.~\cite{Pelaez:2019eqa}, but when evaluating the corresponding K-matrices we use the iso-symmetric mass $M_\pi=135.0 \, {\rm MeV}$.

\subsection{Chiral perturbation theory}
\label{app:chpt}

Following ref.~\cite{Gasser:1983yg}, the $\pi \pi \to \pi \pi$ scattering amplitudes, across all isospin channels, can be written in terms of the amplitude $A(s, t, u)$ defined via
\begin{equation}
M_{ik \to \ell m}(s,t,u) = \delta_{ik} \delta_{\ell m} A(s,t,u)+ \delta_{i \ell} \delta_{km} A(t,s,u) + \delta_{im} \delta_{k \ell} A(u,t,s)\,,
\end{equation}
where $M_{ik \to \ell m}(s,t,u)$, in turn, can be defined in terms of the time ordered four-point function of pions in the real basis $\pi_k(x) = \pi_k(x)^\dagger$ (with $k=1,2,3$). The exact definition reads
\begin{multline}
M_{ik \to \ell m}(s,t,u) = i^3 \lim_{p_i^2 \to M_\pi^2} \bigg [ \prod_{i=1}^4 (M_\pi^2 - p_i^2) \bigg ] \times \\ \int d^4 x_1 d^4 x_2 d^4 x_3 \, e^{i (p_1 \cdot x_1 + p_2 \cdot x_2 + p_3 \cdot x_3 + p_4 \cdot x_4 )} \, \langle 0 \vert {\rm{T}} \big \{\pi_i(x_1) \, \pi_k(x_2)\, \pi_\ell(x_3)\, \pi_m(x_4) \big \} \vert 0 \rangle
\,,
\end{multline}
where the Mandelstam variables are defined as
\begin{equation}
s = (p_1+p_2)^2 \,, \qquad t = (p_1 + p_3)^2 \,, \qquad u = (p_1 + p_4)^2 \,.
\end{equation}
Here we use the mostly minus metric such that $p^2 = (p^0)^2 - \boldsymbol p^2$. Note also that, for the case of on-shell momenta $p_i^2 = M_\pi^2$, one has
\begin{equation}
s + t + u = 4 M_\pi^2 \,.
\end{equation}
The real pion fields apeparing in the four-point function are related to complex pions via
\begin{align}
\begin{pmatrix}
\pi_0(x) & \sqrt 2 \pi^+(x) \\ \sqrt2 \pi^-(x) & -\pi_0(x)
\end{pmatrix} = \sum_{k=1}^3 \pi_k(x) \tau_k \,,
\end{align}
where $\tau_k$ are the usual Pauli matrices.

The definite isospin amplitudes are then given by
\begin{align}
\mathcal M^{[I=0]}(s, t) & = 3 A(s,t,u) + A(t,u,s) + A(u,s,t) \,, \\
\mathcal M^{[I=1]}(s, t) & = A(t,u,s) - A(u,s,t) \,, \\
\mathcal M^{[I=2]}(s, t) & = A(t,u,s) + A(u,s,t) \,,
\end{align}
and the projection of these into partial waves follows from
\begin{equation}
\mathcal M^{[I]}(s, t) = \sum_{\ell=0}^\infty (2 \ell+1) \mathcal M^{[I]}_{\ell}(s) P_{\ell}(\cos \theta) \,,
\end{equation}
with
\begin{equation}
s = 4 (M_\pi^2 + p^2) \,, \qquad t = - 2 p^2 (1 -
\cos \theta) \,.
\end{equation}

Through one-loop in SU(2) chiral perturbation theory, the generating amplitude $A(s, t, u)$ can be written as
\begin{equation}
A(s,t,u) = \frac{s - M^2}{F^2} + B(s,t,u) + C(s,t,u) + \mathcal O(p^6) \,,
\end{equation}
where $B(s,t,u)$ and $C(s,t,u)$ give the logarithmic and analytic polynomial contributions, respectively, arising at next to leading order.

These are given by
\begin{align}
\begin{split}
B\left(s, t, u \right) =& \left(6 F^4\right)^{-1}\bigg [3\left(s^2-M^4\right) \bar{J}(s)\\
& +\left\{t(t-u)-2 M^2 t+4 M^2 u-2 M^4\right\} \bar{J}(t) \\
& +\left\{u(u-t)-2 M^2 u+4 M^2 t-2 M^4\right\} \bar{J}(u)\bigg] \,,
\end{split}\\
\begin{split}
C(s, t, u) =&\left(96 \pi^2 F^4\right)^{-1}\bigg[2\left(\bar{l}_1-\frac{4}{3}\right)\left(s-2 M^2\right)^2\\
& +\left(\bar{l}_2-\frac{5}{6}\right)\left\{s^2+(t-u)^2\right\}-12 M^2 s+15 M^4\bigg] \,,
\end{split}
\end{align}
where:
\begin{align}
\bar{J}\left(q^2\right) =\frac{1}{16 \pi^2}\left\{\sigma \ln \frac{\sigma-1}{\sigma+1}+2\right\},\quad
\sigma =\left(1-\frac{4 M^2}{q^2}\right)^{1 / 2} \,,
\end{align}
and the constants are taken from ref.~\cite{Nebreda:2012ve}:
\begin{align}
F = 92.2\, \text{MeV} , \qquad
\bar{l}_1=-1.5\pm 0.5,\qquad\bar{l}_2=5.3\pm0.7 \,.
\end{align}

As stressed in the main text, we use the ChPT prediction for all channels besides the scalar-isoscalar, for which we use the dispersive result summarized in appendix~\ref{app:sigma}. For all ChPT applications we take the leading-order non-zero contribution. In particular, for the isospin-two $S$-wave we take only the $(s-M^2)/F^2$ expression while for $D$- and $G$-wave we require the contributions from $B(s,t,u)$ and $C(s,t,u)$. In all cases we set $M = M_\pi =135 \, \text{MeV}$, which is valid to leading order in the expansion.

\bibliographystyle{JHEP}
\bibliography{refs.bib}
\end{document}